\documentclass[twocolumn,superscriptaddress,pra]{revtex4-1}%
\usepackage{amsfonts}
\usepackage{amsmath}
\usepackage{amssymb}
\usepackage{graphicx}%
\renewcommand{\vec}[1]{\mathbf{#1}}
\newcommand{\ket}[1]{\left|{#1}\right\rangle}
\newcommand{\tm}[1]{\textrm{#1}}

\begin{document}

\title{Two-state Bogoliubov theory of a molecular Bose gas}

\author{Brandon M.~Peden}
\email[Contact author: ]{brandon.peden@wwu.edu}
\affiliation{Department of Physics and Astronomy, Western Washington University, Bellingham, Washington 98225, USA}
\author{Ryan M.~Wilson}
\affiliation{Department of Physics, The United States Naval Academy, Annapolis, Maryland 21402, USA}
\author{Maverick L.~McLanahan}
\affiliation{Department of Physics and Astronomy, Western Washington University, Bellingham, Washington 98225, USA}
\author{Jesse Hall}
\affiliation{Department of Physics and Astronomy, Western Washington University, Bellingham, Washington 98225, USA}
\author{Seth T.~Rittenhouse}
\affiliation{Department of Physics and Astronomy, Western Washington University, Bellingham, Washington 98225, USA}
\affiliation{Department of Physics, The United States Naval Academy, Annapolis, Maryland 21402, USA}

\date{\today}

\begin{abstract}

We present an analytic Bogoliubov description of a BEC of polar molecules trapped in a quasi-2D geometry and interacting via internal state-dependent dipole-dipole interactions. We derive the mean-field ground-state energy functional, and we derive analytic expressions for the dispersion relations, Bogoliubov amplitudes, and dynamic structure factors. This method can be applied to any homogeneous, two-component system with linear coupling, and direct, momentum-dependent interactions. The properties of the mean-field ground state, including polarization and stability, are investigated, and we identify three distinct instabilities: a density-wave rotonization that occurs when the gas is fully polarized, a spin-wave rotonization that occurs near zero polarization, and a mixed instability at intermediate fields. These instabilities are clarified by means of the real-space density-density correlation functions, which characterize the spontaneous fluctuations of the ground state, and the momentum-space structure factors, which characterize the response of the system to external perturbations. We find that the gas is susceptible to both density-wave and spin-wave response in the polarized limit but only a spin-wave response in the zero-polarization limit. These results are relevant for experiments with rigid rotor molecules such as RbCs, $\Lambda$-doublet molecules such as ThO that have an anomalously small zero-field splitting, and doublet-$\Sigma$ molecules such as SrF where two low-lying opposite-parity states can be tuned to zero splitting by an external magnetic field.

\end{abstract}
\maketitle

\section{Introduction\label{sec:Introduction}}

The experimental realization of Bose-Einstein condensation~\cite{Anderson1995,Bradley1995,Davis1995} and Fermi degeneracy~\cite{DeMarco99,OHara02} in dilute samples of alkali atoms enabled many new discoveries and advances in the field of ultracold degenerate gases including the demonstration of the crossover from a Bose-Einstein condensate (BEC) to a Bardeen-Cooper-Schrieffer superfluid state~\cite{Greiner03,Bourdel04} and the formation of self-assembled vortex lattices~\cite{Abo01,Schweikhard04}. Additionally, the microscopic ``spin'' degrees of freedom in these atomic systems have been used to explore more unconventional states of quantum matter, such as spin-orbit coupled Bose gases~\cite{Lin11,Galitski13} and high-spin Bose gases~\cite{Kawaguchi12,Stamper13}, which are host to a variety of novel quantum phases and phase transitions~\cite{Stenger98,Schmaljohann04,Sadler06}. In all of these systems, ultracold temperatures have permitted the observation of coherent phenomena in the presence of relatively \emph{weak} interactions.

Currently, promising candidates for realizing strong interactions are diatomic, heteronuclear molecules which can possess large electric dipole moments and interact strongly, even in very dilute molecular samples.  Further, dipole-dipole interactions (ddi) are inherently long-range ($\propto 1/r^3$) and anisotropic~\cite{BCtext}. In recent years, experimental groups have made remarkable progress toward cooling molecular samples to quantum degeneracy~\cite{Doyle95,Bethlem99,Nikolov00,Jochim03,Kerman04,Haimberger04,Sage05,Ni08,Deiglmayr08,Danzl10,Stuhl12,Takekoshi2014a,Molony14,Ospelkaus08,Lang08,Kuznetsova09,Aikawa10}. Thus, ultracold molecules are among the most exciting prospects for future studies of strongly interacting quantum many-body systems~\cite{Carr09,Lemeshko13}.

For a large class of molecules (the ``rigid rotors'' e.g. KRb, RbCs, etc.), the lowest-lying microscopic degrees of freedom are rotational in nature, with characteristic energy splittings on the order of $B \sim 1~\mathrm{GHz}$~\cite{BCtext}.  A number of theoretical proposals have discussed how these rotational levels can be manipulated to behave like ``spins,'' and how the state-dependent dipole-dipole interactions can be tuned (using a combination of DC electric and microwave fields) to emulate a broad class of quantum spin models, and thus to study quantum magnetism in a completely new context~\cite{Micheli06,Sadler06,Yao13,Manmana13,Hazzard13,Barnett06,Gorshkov11,Gorshkov11PRA,Kuns11,Herrera11,Xiang12,Lemeshko12,Yan13}. Other molecules, such as $\Lambda$-doublet (e.g. ThO, TiO) or doublet-$\Sigma$ (e.g. SrF) molecules, possess a set of low-lying opposite-parity electronic states with anomalously small energy separations $\sim 10 \, \mathrm{kHz}$~\cite{BCtext,Bohn09,Vutha11}. Recently, a sample of $\Lambda$-doublet OH molecules was Stark-decelerated and evaporatively cooled to temperatures $\lesssim 5 \, \mathrm{mK}$, approaching the quantum degenerate regime~\cite{Stuhl12}.  Unlike the rigid-rotors, the ground state of these molecules forms an effective spin-1/2 manifold, which is energetically far-removed from the higher-lying rotational states.  Even in a very dilute sample, the dipole-dipole interaction energy can approach the doublet splitting, resulting in interesting dielectric properties~\cite{Wilson14}.

Motivated by the experimental progress in the cooling and trapping of heteronuclear polar molecules, there has been a great deal of theoretical interest in understanding the role that strong dipole-dipole interactions play in BECs of molecules that possess spatial degrees of freedom.  Many predictions have been made, including the emergence of a roton-maxon quasiparticle spectrum~\cite{Santos03,Ronen06}, anisotropic superfluid flow~\cite{Ticknor11}, and structured vortex excitations~\cite{Yi06,Wilson09}.  However, very little work has been done to understand the role that the microscopic molecular structure plays in such systems.

In this paper, we present a robust theoretical description of bosonic molecules cooled to quantum degeneracy in which the microscopic nature of the molecules plays an important role. We investigate the mean-field ground state and mesoscopic structure of low-energy excitations by way of Bogoliubov-de Gennes perturbation theory. We present a general, analytic procedure for diagonalizing the fluctuation Hamiltonian, which results in analytic expressions for both the dispersion relations and Bogoliubov amplitudes, in terms of which we can calculate important many-body quantities such as two-point correlation functions and response functions. This procedure generalizes the method developed in Ref.~\cite{Tommasini03} for the case of momentum-dependent couplings that arise as a consequence of the long-range nature of the interactions.

Using these methods, we investigate a quasi-2D BEC of polar molecules in the presence of an external electric field that couples two low-lying molecular states where the molecules interact via state-dependent dipole-dipole interactions. We investigate the properties of the mean-field ground state by way of a Gaussian ansatz for the axial wave functions. By carefully investigating the nature of the two-point density-density and spin-spin correlation functions, we arrive at a complete physical picture of the dynamical instabilities that arise at large densities. We identify three distinct mechanisms for these instabilities, and we conclude with a discussion of three different candidate molecules and the associated parameter regimes (zero-field splitting, field strength, and density) to which these results apply.

The paper is organized as follows. In Sec.~\ref{sec:SingleMoleculeTheory}, we present the theory for the internal structure of the molecules, developing a two-state approximation that allows for a unified treatment of a variety of different molecules. In Sec.~\ref{sec:ManyBodyTheory}, we present the many-body treatment of the system where the many-body Hamiltonian, the Bogoliubov-de Gennes expansion, and the ground state energy functional in a Gaussian approximation are developed. In addition, we present the full analytic diagonalization of the fluctuation Hamiltonian, and thereby derive analytic expressions for important quantities such static response functions. In Sec.~\ref{sec:GroundStateProperties}, we analyze the ground state energy functional and the behavior of energy and polarization of the mean-field ground state, including important limits. In Sec.~\ref{sec:ManyBodyProperties}, we present analyses of the dispersion relations, depletions, static structure factors, and correlation functions all in the context of understanding the nature of the instabilities that appear for large enough density in the low-, intermediate-, and high-field regimes. In Sec.~\ref{sec:Discussion}, we develop a full physical picture of the instabilities seen these regimes through analysis of the static structure factors and correlation functions. Finally, in Sec.~\ref{sec:Conclusion}, we conclude with a discussion of the implications of these results, including how to experimentally access the behavior using the candidate molecules addressed in Sec.~\ref{sec:SingleMoleculeTheory}.

\section{Single Molecule Theory\label{sec:SingleMoleculeTheory}}

In this paper, we consider a gas of polar molecules interacting via the dipole-dipole interaction in the presence of an external electric field $\vec{E}$. In such systems, the net polarization $\vec{P}$ of the gas and the external field together induce a dipole moment $\vec{d}$ in a particular molecule, which in turn modifies the overall polarization. In a semi-classical treatment~\cite{Wilson12}, we solve for both $\vec{d}$ and $\vec{P}$ self-consistently. Here, we are explicitly interested in the role that the microscopic molecular structure plays in determining the many-body behavior of a quantum degenerate gas of polar molecules. We build in a microscopic, quantum mechanical description of polarizability in molecular systems by including two low-lying opposite-parity states of the molecule that are coupled by an external field. This two-state approximation is general enough to provide a unified treatment of a wide class of molecules, including $\Lambda$-doublets, doublet-$\Sigma $'s, and rigid rotor molecules. We note that this description provides a unified picture of both dielectric physics---by building in the microscopic description of molecule polarizability---and of spin-$\frac{1}{2}$ systems with long-range interactions.

This description takes the form of a two-state molecular Hamiltonian, given by
\begin{align}
H_{\mathrm{mol}}=h_{0}\sigma_{z}+h_{c}\sigma_{x},
\label{eqn:molecularHamiltonian}
\end{align}
where $h_{0}  = dE$, $h_{c}  = \Delta/2$, $d$ is the effective dipole moment of the molecule in the strong-field limit, $E$ is the strength of the applied electric field, and $\Delta$ is the zero-field splitting between two low-lying molecular states, the nature of which we will discuss below in the context of specific molecules. This Hamiltonian is written in the basis $\{\ket{\uparrow},\ket{\downarrow}\}$ that diagonalizes the ${d}_0$, where ${d}_0$ is the matrix of the dipole operator that lies along the molecular axis restricted to the lowest two molecular eigenstates. In this case,
\begin{equation}
\hat{d}\rightarrow\left[
\begin{array}
[c]{cc}%
d_{\downarrow} & 0\\
0 & d_{\uparrow}%
\end{array}
\right]   .
\end{equation}
We interpret the basis states in the basis as dipole states show dipole moments align ($\ket{\uparrow}$) or anti-align ($\ket{\downarrow}$) with the external field.

This description is convenient for multiple reasons.  It provides a clear physical picture of the emergent physics, and it eliminates exchange interactions between molecules in the many-body Hamiltonian, enabling a fully analytic solution of the problem within the Bogoliubov de-Gennes framework.
In addition, this allows for a unified many-body description of a wide class of dipolar BEC's.

We have identified three classes of polar molecules relevant to modern experiments that are good candidates for experimentally realizing the results in this paper. These candidate classes are the rigid rotor molecules, $\Lambda$-doublets, and doublet-$\Sigma$s. The specific candidate molecule in the class of rigid rotors is RbCs, which has been cooled by means of both STIRAP~\cite{Takekoshi2014a, Ni08} and photoassociation~\cite{Ji2012}. In the class of $\Lambda$-doublets, we consider ThO, which is a candidate for eEDM searches~\cite{Vutha11}. In the class of doublet-$\Sigma$'s, we consider SrF, which has been laser cooled~\cite{Shuman2010, Barry2012} and is a candidate for realizing magnetic Frenkel excitons in an optical lattice filled with such molecules~\cite{Perez-Rios2010}.

The following discussions of rigid rotor and $\Lambda$-doublet molecules closely follow the discussions in Ref.~\cite{Bohn09}. The discussion of the doublet-$\Sigma$ molecules closely follows the discussion in Ref.~\cite{Perez-Rios2010}. In Sec.~\ref{sec:Discussion}, we give a detailed accounting of the parameter regimes relevant to realizing the results discussed later in this paper for a subset of the molecules described in the following subsections.

\subsection{Rigid rotors}

The Hamiltonian of a rotating molecule in the presence of an external electric
field $\vec{E}$ is given by%
\begin{equation}
\hat{H}_{\mathrm{mol}}=B\hat{J}^{2}-\hat{\vec{d}}\cdot\vec{E},
\end{equation}
where $B$ is the rotational constant, $\vec{J}$ is the total spatial angular
momentum of the molecule, and $\hat{\vec{d}}$ is the dipole moment
operator in the body-fixed frame.
In the basis $\left\{  \left\vert J,M\right\rangle \right\}  $ of
eigenstates of $\hat{J}^{2}$ and $\hat{J}_{z}$, the Hamiltonian is given by
\begin{align}
\hat{H}_{\mathrm{mol}}  &  =\frac{B}{2}\sum_{J,M}J\left(  J+1\right)
\left\vert J,M\right\rangle \left\langle J,M\right\vert \nonumber\\
&  \quad\mbox{}-\sum_{q}E_{q}\sum_{JJ^{\prime}MM^{\prime}}\langle J,M|\hat
{d}_{q}|J^{\prime},M^{\prime}\rangle|J,M\rangle\langle J^{\prime},M^{\prime}|,
\end{align}
and the matrix elements of the dipole operator components $\hat{d}_{q}$ can be
compactly expressed in terms of $3j$ symbols as
\begin{align}
\langle J,M\vert\hat{d}_{q}\vert J^{\prime},M^{\prime}\rangle &  =d\left(
-1\right)  ^{M+q}\sqrt{\left(  2J+1\right)  \left(  2J^{\prime}+1\right)
}\nonumber\\
&  \quad\mbox{}\times\left(
\begin{array}
[c]{ccc}%
J & 1 & J^{\prime}\\
-M & q & M^{\prime}%
\end{array}
\right)  \left(
\begin{array}
[c]{ccc}%
J & 1 & J^{\prime}\\
0 & 0 & 0
\end{array}
\right)  . \label{eqn:dipolematrixelements}%
\end{align}
The $3j$ symbols enforce the selection rules $q+M^{\prime} = M$ and $J+1+J^{\prime}$ are even.
Assuming that the external field is homogeneous and point along the lab-frame $z$-axis, the only term that survives is the $q=0$ term.

\begin{figure}[tb]
\includegraphics[width=86mm]{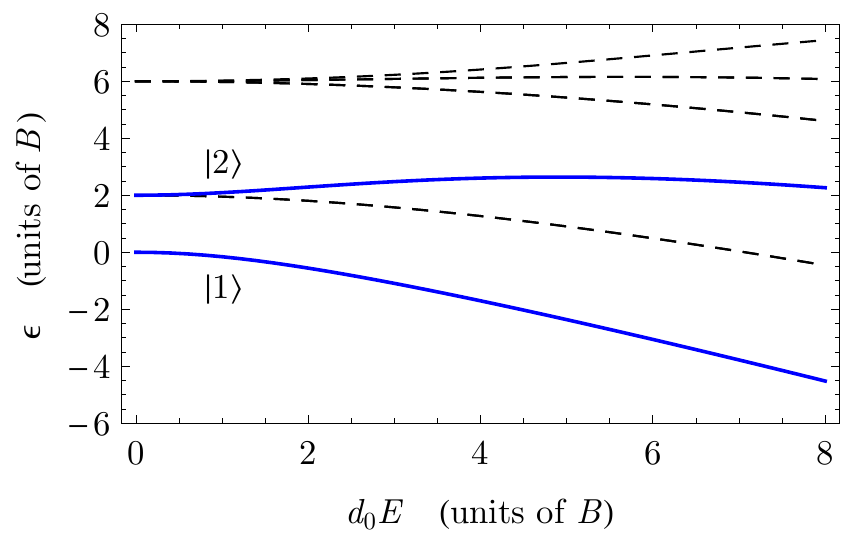}
\caption{(Color online.) The first nine eigenenergies of the Hamiltonian of the rigid rotor molecule in the presence of an external electric field. We restrict to the lowest two $m=0$ states $|1\rangle$ and $|2\rangle$ (solid blue) coupled by the external field.}
\label{fig:RotorEnergies}%
\end{figure}

The external electric field acts to mix angular momentum states according to
Eq.~(\ref{eqn:dipolematrixelements}). By diagonalizing $\hat{H}_{\mathrm{mol}%
}$, we can systematically include the mixing in of higher rotational states
while still treating the system in a two-state approximation. In Fig.~\ref{fig:RotorEnergies}, we have plotted the lowest nine eigenergies of
$\hat{H}_{\mathrm{mol}}$. We keep the lowest eigenstates $\left\vert
1\right\rangle $ and $\left\vert 2\right\rangle $ of $\hat{H}_{\mathrm{mol}}$,
which are adiabatically connected at zero field to the states $\left\vert
00\right\rangle $ and $\left\vert 10\right\rangle $, respectively. In this
truncated basis, the molecular Hamiltonian takes the diagonal form,%
\begin{equation}
\hat{H}_{\mathrm{mol}}=\frac{\epsilon}{2}\left\vert 2\right\rangle
\left\langle 2\right\vert -\frac{\epsilon}{2}\left\vert 1\right\rangle
\left\langle 1\right\vert ,
\end{equation}
where the splitting $\epsilon$ is a function of the field strength $E$; we have subtracted off a constant, field-independent offset. An effective dipole operator can be written in this truncated basis as
\begin{equation}
\hat{d}_{0}^{(\textrm{eff})}=\sum_{\lambda\lambda^{\prime}}d_{\lambda^{\prime}\lambda
}\left\vert \lambda^{\prime}\right\rangle \left\langle \lambda\right\vert ,
\end{equation}
where the matrix elements $d_{\lambda^{\prime}\lambda}$ in the basis $\{\left\vert1\right\rangle,\left\vert 2\right\rangle \}$ are calculated  by constructing $\hat{d}_0$ in the $\ket{J,M}$ basis using Eq.~(\ref{eqn:dipolematrixelements}), transforming to the eigenbasis of $\hat{H}_{\textrm{mol}}$, and restricting to the two lowest states $\ket{1}$ and $\ket{2}$.

\begin{figure}[tb]
\includegraphics[width=86mm]{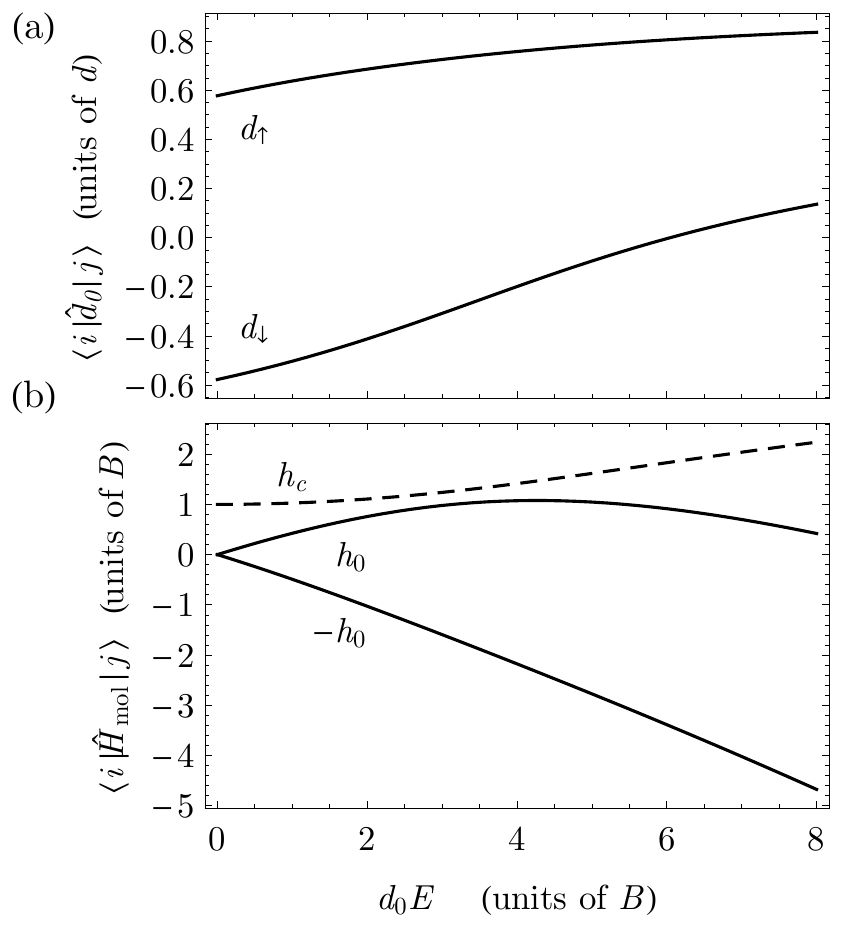}
\caption{Matrix elements of (a) the dipole operator $\hat{d}_{0}$ and (b) the molecular Hamiltonian $\hat{H}_{\text{mol}}$ in the eigenbasis of $\hat{d}_{0}$ restricted to the lowest two eigenstate of $\hat{H}_{\text{mol}}$. For values of $dE\lesssim B$, the dipole moments and the off-diagonal elements of the Hamiltonian are approximately constant, and the diagonal elements of the Hamiltonian grow linearly with the field $E$.}
\label{fig:RotorParameters}
\end{figure}

As noted previously, it is convenient to work in the eigenbasis $\left\{  \left\vert \downarrow\right\rangle ,\left\vert \uparrow\right\rangle \right\} $ of the effective dipole moment operator, in which case the molecular Hamiltonian takes the form of Eq.~(\ref{eqn:molecularHamiltonian}) with
\begin{subequations}
\begin{align}
h_{0} &  =\frac{1}{2}\frac{\epsilon}{\sqrt{\delta^{2}+d_{12}^{2}}}\delta,\\
h_{c} &  =\frac{1}{2}\frac{\epsilon}{\sqrt{\delta^{2}+d_{12}^{2}}}d_{12},\\
\delta &  =\frac{d_{11}-d_{22}}{2}.
\end{align}
\end{subequations}
The dipole moments are given by
\begin{subequations}
\begin{align}
d_{\uparrow} &  =\frac{d_{11}+d_{22}}{2}+\sqrt{\delta^{2}+d_{12}^{2}},\\
d_{\downarrow} &  =\frac{d_{11}+d_{22}}{2}-\sqrt{\delta^{2}+d_{12}^{2}}.
\end{align}
\end{subequations}
In Fig.~\ref{fig:RotorParameters}, we have plotted the matrix elements of both
$\hat{d}_{0}$ and $\hat{H}_{\mathrm{mol}}$ as a function of $d_{0}E$. For
$d_{0}E\lesssim 2B$, $h_{c}$ and $d_{\kappa}$ are constant, and $h_{0}$ varies
linearly with the external field. We therefore interpret the state $\left\vert
\uparrow\right\rangle $ $\left(  \left\vert \downarrow\right\rangle \right)  $
as a molecular state with dipole moment $d_{\uparrow}$ $\left(  d_{\downarrow
}=-d_{\uparrow}\right)  $ that is aligned (anti-aligned) with the external
field. We interpret $2h_{c}$ as a zero-field splitting of the molecule, and
the energies $\pm h_{0}$ of the dipole states display a linear Stark shift.
This interpretation requires that we work in the low-field limit where the
effects of higher-lying rotational states are minimized.

\subsection{$\Lambda$-doublets}

In a certain class of molecules, there exist two low-lying states $\left\vert
1\right\rangle $ and $\left\vert 2\right\rangle $ of opposite parity whose
splitting $\Delta$ is much smaller than the rotational spitting $B$. These two
states are said to comprise a $\Lambda$-doublet~\cite{Bohn09}. The
Hamiltonian for such molecules is given by
\begin{equation}
\label{eqn:MoleculeParityHamiltonian}
\hat{H}_{\mathrm{mol}}=\frac{\Delta}{2}\left( \left\vert 2\right\rangle \left\langle
2\right\vert -\left\vert 1\right\rangle \left\langle
1\right\vert  \right) -d_{0}E \left( \left\vert 1\right\rangle \left\langle 2\right\vert
 + \left\vert 2\right\rangle \left\langle 1\right\vert \right) ,
\end{equation}
where $\vec{E}=E\hat{\mathbf{z}}$ is an external electric field oriented along
the molecular axis, $\Delta\ll B$ is the zero-field splitting of the molecule,
and $d_{0}$ is a function of the total electronic and rotational angular
momentum quantum numbers of the states $\left\vert 2\right\rangle $ and
$\left\vert 1\right\rangle $.

In the basis $\{\ket{\uparrow},\ket{\downarrow}\}$ that diagonalizes $\hat{d}_0$, the molecular Hamiltonian again takes the form of Eq.~(\ref{eqn:molecularHamiltonian}), where $h_{0} = d_{0}E$ and $h_{c} = \Delta/2$, and the dipole moments are  given by $d_{\uparrow} = d_0 = -d_{\downarrow}$. We again interpret these two states as dipole states that either align or anti-align with the external field. These states are the strong-field states of the Hamiltonian, and since $\Delta\ll B$, we can interpolate between the weak-field limit $d_{0}E\ll\Delta$ and the strong field limits $d_{0}E\gg\Delta$ while still neglecting the effects of rotations of the molecule.

\subsection{Doublet-$\Sigma$}

Following the discussion in Ref.~\cite{Perez-Rios2010}, the Hamiltonian for a $^{2}\Sigma$ molecule in the presence of external electric and magnetic fields can be written as
\begin{equation}
\hat{H}_{\mathrm{mol}}=\hat{H}_{\mathrm{ro-vib}}+\gamma_{SR}\mathbf{\hat
{S}\cdot\hat{N}}-\mathbf{\hat{d}}\cdot\mathbf{E}+2\mu_{\mathrm{B}}%
\mathbf{\hat{S}\cdot\hat{B},}%
\end{equation}
where the first term includes both vibrational and rotational terms,
$\mathbf{\hat{S}}$ is the molecular spin, $\mathbf{\hat{N}}$ is the rotational
angular momentum, $\mathbf{\hat{d}}$ is the molecular dipole moment and
$\mu_{\mathrm{B}}$ is the Bohr magneton. We assume that the molecules are in
their vibrational ground states and can be approximated as rigid rotors.
 It is shown in Ref.~\cite{Perez-Rios2010} that there are two
low-lying states $\left\vert 1\right\rangle$ ($N=0$) and $\left\vert 2\right\rangle$ ($N=1$)
of opposite parity that can be tuned to zero splitting via the external
magnetic field. The electric field is the only term in the Hamiltonian that
couples opposite-parity eigenstates, so this crossing is exact at zero-field.
Restricting our attention to only these two states, we can write the
Hamiltonian in the form of Eq.~(\ref{eqn:MoleculeParityHamiltonian}),
where $\Delta$ can now be tuned by an external magnetic field to be
much smaller than the rotational splitting. We again work
in the basis where the dipole moment operator $\hat{d}_{0}$ is diagonal, and
everything carries over from the $\Lambda$-doublet section.

\section{Many-Body Hamiltonian and Bogoliubov-de Gennes Analysis\label{sec:ManyBodyTheory}}

In this section, we present the general many-body treatment of a molecular BEC interacting via electric dipole interactions in the case where the molecules can be treated in a two-state approximation, as discussed in Sec.~\ref{sec:SingleMoleculeTheory}. Using a Bogoliubov-de Gennes analysis, we derive both the ground-state energy functional $K_0$ and the second-order fluctuation Hamiltonian $\hat{K}_2$ in the grand-canonical ensemble via a Gaussian ansatz for the axial (trap-axis) wave functions. We analytically diagonalize the fluctuation Hamiltonian. This procedure results in analytic expressions for the low-energy dispersion relations and Bogoliubov amplitudes, in terms of which we can write important many-body properties such as the quantum depletion and static structure factors.

\subsection{Many-body Hamiltonian in the dipole basis}

The full many-body Hamiltonian is given by%
\begin{align}
\hat{H}  &  = \hat{H}_{0}+\hat{H}_{\mathrm{int}}\nonumber\\
&  = \int d^{3}r\hat{\Psi}^{\dagger}\left(  \vec{r}\right)  \left(
H_{\textrm{CM}}\left(  \vec{r}\right)  \hat{1}+\hat{H}_{\mathrm{mol}}\right)  \hat{\Psi
}\left(  \vec{r}\right) \nonumber\\
&  \quad\mbox{}+\frac{1}{2}\int d^{3}r\int d^{3}r^{\prime}\hat{\Psi}^{\dagger
}\left(  \vec{r}\right)  \hat{\Psi}^{\dagger}\left(  \vec{r}^{\prime}\right)
\nonumber\\
&  \quad\mbox{}\times\hat{U}\left(  \vec{r}-\vec{r}^{\prime}\right)  \hat
{\Psi}\left(  \vec{r}\right)  \hat{\Psi}\left(  \vec{r}^{\prime}\right)  ,
\label{eqn:Hamiltonian}%
\end{align}
where $H_{\textrm{CM}}\left(  \vec{r}\right)  $ is the single-particle Hamiltonian for
the center-of-mass motion of the molecule, and%
\begin{equation}
\hat{U}\left(  \vec{r}_{1}-\vec{r}_{2}\right)  =\hat{d}^{(1)}_{0}\hat{d}^{(2)}_{0}%
\frac{1-3\cos^{2}\theta_{\vec{r}_{1}-\vec{r}_{2}}}{\left\vert \vec{r}_{1}%
-\vec{r}_{2}\right\vert ^{3}},
\end{equation}
where we have assumed that the induced dipoles lie along the $z$-axis.

Expanding the field operator as a two-component spinor
\begin{equation}
\hat{\Psi}\left(  \vec{r}\right)  =\sum_{\kappa\in\{\uparrow,\downarrow\}}\hat{\psi}_{\kappa
}\left(  \vec{r}\right)  \left\vert \kappa\right\rangle ,
\end{equation}
the interaction Hamiltonian becomes%
\begin{align}
\hat{H}_{\mathrm{int}}  &  =\frac{1}{2}\sum_{\kappa,\kappa^{\prime}}d_{\kappa
}d_{\kappa^{\prime}}\int d^{3}r\int d^{3}r^{\prime}V\left(  \vec{r}-\vec
{r}^{\prime}\right) \nonumber\\
&  \quad\mbox{}\times\hat{\psi}_{\kappa}^{\dagger}\left(  \vec{r}\right)
\hat{\psi}_{\kappa^{\prime}}^{\dagger}\left(  \vec{r}^{\prime}\right)
\hat{\psi}_{\kappa^{\prime}}\left(  \vec{r}^{\prime}\right)  \hat{\psi
}_{\kappa}\left(  \vec{r}\right)  ,
\label{eqn:InteractionHamiltonian}
\end{align}
where%
\begin{equation}
V\left(  \vec{r}-\vec{r}^{\prime}\right)  =\frac{1-3\cos^{2}\theta_{\vec
{r}-\vec{r}^{\prime}}}{\left\vert \vec{r}-\vec{r}^{\prime}\right\vert ^{3}}~.
\end{equation}
The rotation to the single-molecule strong-field basis removes any exchange interactions in the Hamiltonian, and we are left with only direct interaction terms.  

The remaining terms in the Hamiltonian (Eq.(\ref{eqn:Hamiltonian})) can be
expressed in terms of the field operators $\hat{\psi}_{\kappa}\left(  \vec{r}\right)  $, and the result is
\begin{equation}
\label{eqn:GeneralHamiltonian}
\hat{K}=\hat{H}_{0}-\sum_{\kappa}\mu_{\kappa}\hat{N}_{\kappa}+\hat
{H}_{\mathrm{lin}}+\hat{H}_{\mathrm{int}},
\end{equation}
where $\hat{H}_{0}$, given by
\begin{align}
\hat{H}_{0}  &  =\sum_{\kappa}\int d^{3}%
r\hat{\psi}_{\kappa}^{\dagger}\left(  \vec{r}\right)  \left(  h_{0}\left(
\vec{r}\right)  +h_{0}\left(  \delta_{\kappa,\downarrow}-\delta_{\kappa
,\uparrow}\right)  \right)  \hat{\psi}_{\kappa}\left(  \vec{r}\right),
\end{align}
is the single-molecule Hamiltonian, and $\hat{H}_{\mathrm{lin}}$, given by
\begin{align}
\hat{H}_{\mathrm{lin}}  &  =h_{c}\int d^{3}r\left(  \hat{\psi}_{\uparrow
}^{\dagger}\left(  \vec{r}\right)  \hat{\psi}_{\downarrow}\left(  \vec
{r}\right)  +\hat{\psi}_{\downarrow}^{\dagger}\left(  \vec{r}\right)
\hat{\psi}_{\uparrow}\left(  \vec{r}\right)  \right),
\end{align}
is the Hamiltonian for the linear coupling between dipole states that arises as a consequence of the zero-field splitting.
We  introduced chemical potentials $\mu_{\kappa}$ to work in the
grand-canonical ensemble, and $\hat{N}_{\kappa}$, given by%
\begin{equation}
\hat{N}_{\kappa}=\int d^{3}r\hat{\psi}_{\kappa}^{\dagger}\left(  \vec
{r}\right)  \hat{\psi}_{\kappa}\left(  \vec{r}\right)  ,
\end{equation}
which is the number operator for the internal state $\kappa$. We consider a gas of polar molecules harmonically trapped in quasi-2D, in which case the Hamiltonian is given in cylindrical coordinates by
\begin{subequations}
\begin{align}
H_{\textrm{CM}}\left(  \vec{r}\right)   &  =-\frac{\hbar^{2}}{2m}{\nabla}_{\rho}%
^{2}+H_{\textrm{CM}}\left(  z\right),
\end{align}
where
\begin{align}
H_{\textrm{CM}}\left(  z\right)   &  =-\frac{\hbar^{2}}{2m}{\nabla}_{z}^{2}+\frac{1}%
{2}m\omega^{2}z^{2}.
\end{align}
\end{subequations}

\subsection{Bogoliubov Theory}

Here, we perform a Bogoliubov-de Gennes analysis of the full many-body Hamiltonian derived above. When the axial trapping is sufficiently tight, we can expand the field operator $\hat{\psi}_{\kappa}\left(  \vec {r}\right)$ as the product of an axial wave function and a field operator for the in-plane motion. We further expand the field operator as a sum of condensate and fluctuation terms, yielding
\begin{align}
\hat{\psi}_{\lambda}\left(  \vec{x}\right)
&  =\frac{f_{\lambda}\left(  z\right)  }{\sqrt{A}}\left(  e^{i\theta_{\lambda
}}\sqrt{N_{\lambda}}+\sum_{\vec{k}\neq0}e^{i\vec{k}\cdot{\boldsymbol\rho}}
\hat{a}_{\vec{k},\lambda}\right)  , \label{eqn:PlaneWaveExpansion}%
\end{align}
where $N_{\lambda}$ is the number of particles occupying molecular state $\ket{\lambda}$, $\vec{k}$ is an in-plane wave vector, $f_{\lambda}\left(z\right)$ is a normalized, state-dependent axial wave function, and $A$ is the in-plane area of the system. Neglecting the fourth-order terms in the expansion, Eq.~(\ref{eqn:GeneralHamiltonian}) becomes
\begin{align}
\hat{K} \approx {K}_0 + \hat{K_2}~,
\end{align}
where ${K}_0$, given by
\begin{align}
{K}_{0}  &  =\sum_{\kappa}\left(  \epsilon_{\kappa}-\mu_{\kappa}%
+h_{0}\left(  \delta_{\kappa,\downarrow}-\delta_{\kappa,\uparrow}\right)
\right)  N_{\kappa}\nonumber\\
&  \quad\mbox{}+\sqrt{N_{\uparrow}N_{\downarrow}}\left(  e^{-i\left(
\theta_{\uparrow}-\theta_{\downarrow}\right)  }\alpha+c.c.\right) \nonumber\\
&  \quad\mbox{}+\frac{1}{2}\sum_{\kappa,\kappa^{\prime}}N_{\kappa}%
N_{\kappa^{\prime}}\frac{l^{2}}{A}\lambda_{\kappa\kappa^{\prime}},
\label{eqn:GroundStateEnergy}%
\end{align}
is the ground-state energy functional, and $\hat{K}_2$, given by
\begin{widetext}
\begin{align}
\hat{K}_{2}  &  =-\frac{1}{2}\sum_{\vec{k}\neq0}\sum_{\kappa}N_{\kappa}%
\frac{l^{2}}{A}\lambda_{\vec{k},\kappa\kappa}
\nonumber\\
&  \quad\mbox{}+
\sum_{\vec{k}\neq0}\sum_{\kappa}
\left(
	\frac{\hbar^{2}k^{2}}{2m} +
	\epsilon_{\kappa}-\mu_{\kappa}+h_{0}\left(  \delta_{\kappa,\downarrow}-\delta_{\kappa,\uparrow}\right)+
	\sum_{\kappa^{\prime}}N_{\kappa^{\prime}}\frac{l^{2}}{A}\lambda_{\kappa\kappa^{\prime}}
\right)
\hat{a}_{\vec{k},\kappa}^{\dagger}\hat{a}_{\vec{k},\kappa}
+\sum_{\vec{k}\neq0}\alpha\hat{a}_{\vec{k},\uparrow}^{\dagger}\hat{a}_{\vec{k},\downarrow}+H.c.
\nonumber\\
&  \quad\mbox{}+\sum_{\vec{k}\neq0}\sum_{\kappa,\kappa^{\prime}}%
\sqrt{N_{\kappa}N_{\kappa^{\prime}}}\frac{l^{2}}{A}\frac{\lambda_{\vec
{k},\kappa\kappa^{\prime}}}{2}
\left(  e^{i\theta_{\kappa}}\hat{a}_{\vec{k},\kappa
}^{\dagger}+e^{-i\theta_{\kappa}}\hat{a}_{-\vec{k},\kappa}\right)
\left(  e^{i\theta_{\kappa^{\prime}}}\hat{a}_{-\vec
{k},\kappa^{\prime}}^{\dagger}+e^{-i\theta_{\kappa^{\prime}}}\hat{a}_{\vec
{k},\kappa^{\prime}}\right)  ,
\label{eqn:FluctuationHamiltonianGeneral}%
\end{align}
\end{widetext}
is the second-order fluctuation Hamiltonian. These terms are expressed in terms of the single-particle parameters,
\begin{align}
\epsilon_{\kappa}  &  =\int dzf_{\kappa}^{\ast}\left(  z\right)  H_{\textrm{CM}}\left(
z\right)  f_{\kappa}\left(  z\right)  ,\\
\alpha &  =h_{c}\int dzf_{\uparrow}^{\ast}\left(  z\right)  f_{\downarrow
}\left(  z\right)  ,
\end{align}
and the interaction parameters,
\begin{subequations}
\label{eqn:InteractionParametersGeneral}
\begin{align}
\lambda_{\vec{k},\kappa\kappa^{\prime}}  &  =\frac{d_{\kappa^{\prime}%
}d_{\kappa}}{l^{2}}\int d^{3}r\int d^{3}r^{\prime}\left\vert f_{\kappa}\left(
z\right)  \right\vert ^{2}\frac{e^{-i\vec{k}\cdot{\boldsymbol\rho}}}{\sqrt{A}%
}\nonumber\\
& \quad \mbox{}\times V\left(  \vec{r}-\vec{r}^{\prime}\right)  \left\vert
f_{\kappa^{\prime}}\left(  z^{\prime}\right)  \right\vert ^{2}\frac
{e^{i\vec{k}\cdot{\boldsymbol\rho}^{\prime}}}{\sqrt{A}},\\
\lambda_{\kappa\kappa^{\prime}}  &  =\lambda_{\vec{k}=0,\kappa\kappa^{\prime}%
},
\end{align}
\end{subequations}
where $l=\sqrt{\hbar/m\omega}$ is the oscillator length associated with the
axial trapping potential.

These equations are purely general for the case where the particles interact exclusively via direct interactions via a state-independent central potential $V$ (see Eq.~(\ref{eqn:InteractionHamiltonian})). In the case of dipole-dipole interactions between dipoles aligned with the trap-axis, the interaction parameter can be written in the simple form (see Appendix \ref{app:InteractionParameters})
\begin{align}
\lambda_{\vec{k},\kappa\kappa^{\prime}}  &  =\frac{d_{\kappa^{\prime}%
}d_{\kappa}}{l^{2}}\frac{8\pi}{3}\int dz\left\vert f_{\kappa}\left(  z\right)
\right\vert ^{2}\left\vert f_{\kappa^{\prime}}\left(  z\right)  \right\vert
^{2}
\nonumber\\
&  \quad\mbox{}-
\frac{d_{\kappa}d_{\kappa^{\prime}}}{l^{2}}2\pi k
\int dzdz^{\prime}e^{-k\left\vert z-z^{\prime}\right\vert }
\nonumber\\
&  \quad\mbox{}\times
\left\vert f_{\kappa}\left(  z\right) \right\vert^{2} \left\vert f_{\kappa^{\prime}}\left(  z^{\prime}\right)  \right\vert^{2}.
\end{align}

In order to calculate the parameters, we need the axial wave functions $f_{\kappa}$. We can minimize the ground state energy with respect to $\theta$ and the population-normalized axial wave functions $F_{\kappa}\left(  z\right)  =\sqrt{N_{\kappa}/N}f_{\kappa}\left(  z\right)$, and enforce the normalization condition $1=\sum_{\kappa}\int dz^{\prime}\left\vert F_{\kappa}\left(  z^{\prime}\right) \right\vert ^{2}$. Extremizing with respect to $\theta$ yields $\theta=\pi$. Extremizing with respect to $F_{\kappa}$ results in a set of coupled differential equations, given by
\begin{align}
0  &  = \left(  h_{0}\left(  z\right)  +\left(  \delta_{\kappa,\downarrow
}-\delta_{\kappa,\uparrow}\right)  h_{0}-\mu_{\kappa}\right)  f_{\kappa
}\left(  z\right) \nonumber\\
&  \quad\mbox{}+h_{c}\cos\theta\left(  \delta_{\kappa,\uparrow}\sqrt
{\frac{N_{\downarrow}}{N_{\uparrow}}}f_{\downarrow}\left(  z\right)
+\delta_{\kappa,\downarrow}\sqrt{\frac{N_{\uparrow}}{N_{\downarrow}}%
}f_{\uparrow}\left(  z\right)  \right) \nonumber\\
&  \quad\mbox{}+\frac{8\pi}{3}\sum_{\kappa^{\prime}%
}\frac{N_{\kappa^{\prime}}d_{\kappa}d_{\kappa^{\prime}}}{A}\left\vert
f_{\kappa^{\prime}}\left(  z\right)  \right\vert ^{2}f_{\kappa}\left(
z\right),
\label{eqn:GPEquations}%
\end{align}
which is constrained by both the normalization condition above and the equilibrium condition, $\mu_{\uparrow}=\mu_{\downarrow}$.

Here, we instead employ a Gaussian ansatz for the axial wave functions, given by
\begin{equation}
f_{\kappa}\left(  z\right)  =\frac{1}{\sqrt{l\sqrt{\pi}}}e^{-z^{2}/2l^{2}}.
\label{eqn:GaussianAnsatz}
\end{equation}
This allows us to find analytic expressions for the interaction parameters, and we find that this ansatz results in good qualitative agreement with the results obtained by using the numerical solutions of  Eq.~(\ref{eqn:GPEquations}). The parameters in Eq.~(\ref{eqn:GroundStateEnergy}) can be evaluated analytically, and the ground state energy per particle can be written as
\begin{align}
\frac{K_{0}}{N}  &  =\left(  \frac{\hbar\omega}{2}-\mu_{\uparrow}%
-h_{0}\right)  \nu_{\uparrow}+\left(  \frac{\hbar\omega}{2}-\mu_{\downarrow
}+h_{0}\right)  \nu_{\downarrow}\nonumber\\
&  \quad\mbox{}+2\alpha\cos\theta\sqrt{\nu_{\uparrow}\nu_{\downarrow}%
}\nonumber\\
&  \quad\mbox{}+\frac{1}{2}\nu_{\uparrow}^{2}n\lambda_{\uparrow\uparrow}%
+\frac{1}{2}\nu_{\downarrow}^{2}n\lambda_{\downarrow\downarrow}+\nu_{\uparrow
}\nu_{\downarrow}n\lambda_{\uparrow\downarrow},
\label{eqn:GroundStateEnergyGaussian}
\end{align}
where $\theta=\theta_{\uparrow}-\theta_{\downarrow}$ is the relative phase between the two components, $n=Nl^{2}/A$ is the total 2D areal density scaled by $l^{-2}$, $\nu_{\kappa}=N_{\kappa}/N$ is the relative number of molecules occupying molecular state $\ket{\kappa}$, and the interaction parameter is given by%
\begin{align}
\lambda_{\kappa\kappa^{\prime}}=\frac{4\sqrt{2\pi}}{3}\frac{d_{\kappa
}d_{\kappa^{\prime}}}{l^{3}}.
\end{align}

In order to find the ground state energy, we minimize Eq.~(\ref{eqn:GroundStateEnergyGaussian}) with
respect to $\theta$ and $\nu_{\kappa}$, yielding $\theta=\pi$ and
\begin{subequations}
\label{eqn:chemicalpotentials}%
\begin{align}
\mu_{\uparrow}  &  =\epsilon_{\uparrow}-h_{0}+\alpha\cos\theta\sqrt{\frac
{\nu_{\downarrow}}{\nu_{\uparrow}}}+\sum_{\kappa}n\nu_{\kappa}\lambda
_{\uparrow\kappa},\\
\mu_{\downarrow}  &  =\epsilon_{\downarrow}+h_{0}+\alpha\cos\theta\sqrt
{\frac{\nu_{\uparrow}}{\nu_{\downarrow}}}+\sum_{\kappa}n\nu_{\kappa}%
\lambda_{\downarrow\kappa},
\end{align}
which determines the relative population via the equilibrium condition,
$\mu_{\uparrow}=\mu_{\downarrow}$.

In light of the Gaussian ansatz and energy minimization procedure, the
fluctuation Hamiltonian (Eq.~(\ref{eqn:FluctuationHamiltonianGeneral})) can be
written as%
\end{subequations}
\begin{align}
\hat{K}_{2}  &  =-\frac{1}{2}\sum_{\kappa,\vec{k}\neq0}\nu_{\kappa}%
n\lambda_{\vec{k},\kappa\kappa}\nonumber\\
&  \quad\mbox{}+\sum_{\kappa,\vec{k}\neq0}\left(  \frac{\hbar\omega\left(
kl\right)  ^{2}}{2}-\alpha\cos\theta\sqrt{\frac{\delta_{\kappa,\downarrow}%
\nu_{\uparrow}+\delta_{\kappa,\uparrow}\nu_{\downarrow}}{\nu_{\kappa}}%
}\right)  \hat{a}_{\vec{k},\kappa}^{\dagger}\hat{a}_{\vec{k},\kappa
}\nonumber\\
&  \quad\mbox{}+\alpha\sum_{\vec{k}\neq0}\hat{a}_{\vec{k},\uparrow}^{\dagger
}\hat{a}_{\vec{k},\downarrow}+H.c.\nonumber\\
&  \quad\mbox{}+\sum_{\vec{k}\neq0}\sum_{\kappa,\kappa^{\prime}}\left(
\cos\theta\right)  ^{^{1-\delta_{\kappa\kappa^{\prime}}}}\frac{\sqrt
{\nu_{\kappa}\nu_{\kappa^{\prime}}}n\lambda_{\vec{k},\kappa\kappa^{\prime}}%
}{2}\nonumber\\
&  \quad\mbox{}\times\left(  \hat{a}_{\vec{k},\kappa}^{\dagger}+\hat{a}%
_{-\vec{k},\kappa}\right)  \left(  \hat{a}_{-\vec{k},\kappa^{\prime}}%
^{\dagger}+\hat{a}_{\vec{k},\kappa^{\prime}}\right)  ,
\end{align}
where the interaction parameter is explicitly given by
\begin{subequations}
\label{eqn:InteractionParameterGaussian}
\begin{align}
\lambda_{\vec{k},\kappa\kappa^{\prime}}
&=
\lambda_{\kappa\kappa^{\prime}}F\left(  \frac{kl}{\sqrt{2}}\right) 
\end{align}
where
\begin{equation}
F\left(  x\right) =1-\frac{3\sqrt{\pi}}{2}xe^{x^{2}}\mathrm{erfc}\left(x\right)  ,
\end{equation}
\end{subequations}
and $\mathrm{erfc}(x)$ is the complementary error function.

\subsection{Diagonalization of the fluctuation Hamiltonian}

In Ref.~\cite{Tommasini03}, the Bogoliubov diagonalization procedure was generalized in order to deal with a coherently coupled two-state BEC whose atoms interact via contact interactions. Here, we generalize this procedure for the case of the momentum-dependent couplings (Eq.~(\ref{eqn:InteractionParameterGaussian})) that arise when the particles interact via an anisotropic, long-range interaction. We note that this procedure can be applied to any BEC with a linear coupling term and state-dependent, momentum-dependent interaction couplings, as long as there are only direct interactions (see Eqs.~(\ref{eqn:InteractionParametersGeneral}) and (\ref{eqn:FluctuationHamiltonianGeneral})). The diagonalization procedure results in \emph{analytic} expressions for the Bogoliubov amplitudes, and they can be combined with (if necessary) numerical values of the interaction parameters to yield important quantities such as response functions.

The diagonalization procedure consists of the following steps: (1) a canonical transformation of the plane-wave operators $\hat{a}_{\vec{k},\kappa}$ that removes the linear coupling between the two modes; (2) a transformation to a set of non-Hermitian coordinate operators scaled in such a way that the momentum terms are left invariant under a further rotation of the coordinate operators; (3) a rotation of the coordinate operators that decouples the two components; and (4) a transformation back to a set of bosonic operators in terms of which the fluctuation Hamiltonian is diagonal. Since in-plane center-of-mass momentum is conserved in this system, this procedure is identical for each block of the Hamiltonian corresponding to a particular momentum.

For purposes of clarity, in what follows, we make the replacement $(\downarrow,\uparrow)\to(1,2)$.

The first transformation is given by
\begin{subequations}
\label{eqn:linearCouplingRotation}%
\begin{align}
\hat{a}_{\vec{k},1}  &  =\hat{d}_{\vec{k},1}\cos\eta-\hat{d}_{\vec
{k},2}\sin\eta,\\
\hat{a}_{\vec{k},2}  &  =\hat{d}_{\vec{k},1}\sin\eta+\hat{d}_{\vec
{k},2}\cos\eta,
\end{align}
\end{subequations}
where%
\begin{subequations}
\begin{align}
\cos\eta &  =\sqrt{\nu_{2}}\frac{1-\cos\theta}{2}+\sqrt{\nu_{1}}\frac
{1+\cos\theta}{2},\\
\sin\eta &  =\sqrt{\nu_{1}}\frac{1-\cos\theta}{2}+\sqrt{\nu_{2}}\frac
{1+\cos\theta}{2},
\end{align}
\end{subequations}
and it can be easily shown that the new creation and annihilation operators $\hat{d}_{\vec{k},\sigma}$ and $\hat{d}_{\vec{k},\sigma}^{\dagger}$ satisfy the canonical boson commutation relations. Under this transformation, the Hamiltonian takes the form
\begin{align}
\hat{K}_{2}  &  =
-\frac{1}{2}\sum_{\kappa,\vec{k}\neq0}\nu_{\kappa}%
n\lambda_{\vec{k},\kappa\kappa}
+\sum_{\vec{k}\neq0}\sum_{\sigma}\epsilon_{k,\sigma}\hat
{d}_{\vec{k},\sigma}^{\dagger}\hat{d}_{\vec{k},\sigma}\nonumber\\
&  \quad\mbox{}+\sum_{\vec{k}\neq0}\sum_{\sigma,\sigma^{\prime}}\frac
{\sqrt{\nu_{\sigma}\nu_{\sigma^{\prime}}}n\Lambda_{\vec{k},\sigma
\sigma^{\prime}}}{2}
\nonumber\\
&\quad\mbox{}\times\left(  \hat{d}_{\vec{k},\sigma}^{\dagger}+\hat{d}%
_{-\vec{k},\sigma}\right)  \left(  \hat{d}_{-\vec{k},\sigma^{\prime}}%
^{\dagger}+\hat{d}_{\vec{k},\sigma^{\prime}}\right)
\end{align}
where%
\begin{subequations}
\begin{align}
\epsilon_{k,2}  &  =\frac{\hbar^{2}k^{2}}{2m}-\frac{1+\cos\theta}{2}%
\frac{\alpha}{\sqrt{\nu_{1}\nu_{2}}},\\
\epsilon_{k,1}  &  =\frac{\hbar^{2}k^{2}}{2m}+\frac{1-\cos\theta}{2}%
\frac{\alpha}{\sqrt{\nu_{1}\nu_{2}}},
\end{align}
\end{subequations}
are the single-particle energies, and
\begin{subequations}
\label{eqn:DressedInteractionCouplings}%
\begin{align}
\nu_{1}\Lambda_{\vec{k},11}  &  =\nu_{1}\lambda_{\vec{k},11}\cos^{2}\eta
+\nu_{2}\lambda_{\vec{k},22}\sin^{2}\eta\\
&  \quad\mbox{}+\sqrt{\nu_{1}\nu_{2}}\operatorname{Re}\left(  \lambda_{\vec
{k},12}\right)  \sin2\eta\cos\theta,\nonumber\\
\nu_{2}\Lambda_{\vec{k},22}  &  =\nu_{2}\lambda_{\vec{k},22}\cos^{2}\eta
+\nu_{1}\lambda_{11}\sin^{2}\eta\\
&  \quad\mbox{}-\sqrt{\nu_{1}\nu_{2}}\operatorname{Re}\left(  \lambda_{\vec
{k},12}\right)  \sin2\eta\cos\theta,\nonumber\\
\sqrt{\nu_{1}\nu_{2}}\Lambda_{\vec{k},12}  &  =\frac{\nu_{2}\lambda_{\vec
{k},22}-\nu_{1}\lambda_{\vec{k},11}}{2}\sin2\eta\nonumber\\
&  \quad\mbox{}+\sqrt{\nu_{1}\nu_{2}}\left(  \lambda_{\vec{k},12}\cos^{2}%
\eta-\lambda_{\vec{k},21}\sin^{2}\eta\right)  \cos\theta,
\end{align}
\end{subequations}
are a set of dressed momentum dependent interaction parameters.

We next define a set of non-Hermitian coordinate operators via the transformation,
\begin{subequations}
\begin{align}
\hat{x}_{\vec{k},\sigma}  &  =\sqrt{\frac{\omega_{\vec{k},1}}{\epsilon
_{k,\sigma}}}\frac{\hat{d}_{\vec{k},\sigma}^{\dagger}+\hat{d}_{-\vec{k}%
,\sigma}}{\sqrt{2}},\\
\hat{p}_{\vec{k},\sigma}  &  =\sqrt{\frac{\epsilon_{k,\sigma}}{\omega_{\vec
{k},1}}}\frac{\hat{d}_{\vec{k},\sigma}-\hat{d}_{-\vec{k},\sigma}^{\dagger}%
}{i\sqrt{2}},
\end{align}
\end{subequations}
where%
\begin{equation}
\omega_{\vec{k},\sigma}=\sqrt{\epsilon_{k,\sigma}\left(  \epsilon_{k,\sigma
}+2n\nu_{\sigma}\Lambda_{\vec{k},\sigma\sigma}\right)  },
\end{equation}
are the dispersion relations in the absence of interactions between anti-aligned dipoles. These coordinate operators are non-Hermitian---i.e.~$\hat{x}_{\vec{k},\sigma}^{\dagger}  =\hat{x}_{-\vec{k},\sigma}$ and $\hat{p}_{\vec{k},\sigma}^{\dagger}  =\hat{p}_{-\vec{k},\sigma}$---but they still satisfy the canonical position-momentum commutation relations. Under this transformation, the Hamiltonian is
\begin{align}
\hat{K}_{2}  &  =-\frac{1}{2}\sum_{\vec{k}\neq0}   \sum_{\sigma}\left(%
\epsilon_{k,\sigma}+\nu_{\sigma}n\lambda_{\vec{k},\sigma\sigma}\right) \nonumber\\
&  \quad\mbox{}+\sum_{\vec{k}\neq0}\sum_{\sigma}\frac{\omega_{\vec{k},1}}{2}\hat
{p}_{-\vec{k},\sigma}\hat{p}_{\vec{k},\sigma}\nonumber\\
&  \quad\mbox{}+\sum_{\vec{k}\neq0}\sum_{\sigma}\frac{\omega_{\vec{k},\sigma}^{2}%
}{2\omega_{\vec{k},1}}\hat{x}_{\vec{k},\sigma}\hat{x}_{-\vec{k},\sigma
}\nonumber\\
&  \quad\mbox{}+\sum_{\vec{k}\neq0}\frac{\sqrt{\nu_{1}\nu_{2}\epsilon_{k,1}%
\epsilon_{k,2}}}{\omega_{\vec{k},1}}\left(  \Lambda_{\vec{k},12}\hat{x}%
_{\vec{k},1}\hat{x}_{-\vec{k},2}+\textrm{H.c.}\right)  .
\end{align}

The next step is to decouple modes 1 and 2 by defining a new set of coordinate
operators,%
\begin{subequations}
\begin{align}
\hat{P}_{\vec{k},+}  &  =\sqrt{\frac{\omega_{\vec{k},1}}{\Omega_{\vec{k},+}}%
}\left(  \hat{p}_{\vec{k},1}\cos\gamma_{\vec{k}}+\hat{p}_{\vec{k},2}%
e^{-i\phi_{\vec{k}}}\sin\gamma_{\vec{k}}\right)  ,\\
\hat{P}_{\vec{k},-}  &  =\sqrt{\frac{\omega_{\vec{k},1}}{\Omega_{\vec{k},-}}%
}\left(  \hat{p}_{\vec{k},2}\cos\gamma_{\vec{k}}-\hat{p}_{\vec{k},1}%
e^{i\phi_{\vec{k}}}\sin\gamma_{\vec{k}}\right)  ,\\
\hat{X}_{\vec{k},+}  &  =\sqrt{\frac{\Omega_{\vec{k},+}}{\omega_{\vec{k},1}}%
}\left(  \hat{x}_{\vec{k},1}\cos\gamma_{\vec{k}}+\hat{x}_{\vec{k},2}%
e^{i\phi_{\vec{k}}}\sin\gamma_{\vec{k}}\right)  ,\\
\hat{X}_{\vec{k},-}  &  =\sqrt{\frac{\Omega_{\vec{k},-}}{\omega_{\vec{k},1}}%
}\left( \hat{x}_{\vec{k},2}\cos\gamma_{\vec{k}}-\hat{x}_{\vec{k},1}%
e^{-i\phi_{\vec{k}}}\sin\gamma_{\vec{k}}\right)  ,
\end{align}
where%
\end{subequations}
\begin{align}
\Omega_{\vec{k},\pm}^{2}  &  =\frac{1}{2}\left(  \omega_{\vec{k},1}^{2}%
+\omega_{\vec{k},2}^{2}\right) \nonumber\\
&  \quad\mbox{}\pm\frac{1}{2}\sqrt{\left(  \omega_{\vec{k},1}^{2}-\omega
_{\vec{k},2}^{2}\right)  ^{2}+16\nu_{1}\nu_{2}n^{2}|\Lambda_{\vec{k},12}%
|^{2}\epsilon_{k,1}\epsilon_{k,2}}, \label{eqn:DispersionRelations}%
\end{align}
and the angle functions are given by
\begin{subequations}
\begin{align}
\cos\gamma_{\vec{k}}  &  =\sqrt{\frac{1}{2}\left(  1+\frac{\omega_{\vec{k}%
,1}^{2}-\omega_{\vec{k},2}^{2}}{\Omega_{\vec{k},+}^{2}-\Omega_{\vec{k},-}^{2}%
}\right)  },\\
\sin\gamma_{\vec{k}}  &  =\sqrt{\frac{1}{2}\left(  1-\frac{\omega_{\vec{k}%
,1}^{2}-\omega_{\vec{k},2}^{2}}{\Omega_{\vec{k},+}^{2}-\Omega_{\vec{k},-}^{2}%
}\right)  },\\
\phi_{\vec{k}}  &  =\arg\left(  \Lambda_{\vec{k},12}\right)  .
\end{align}
Again, these operators are not Hermitian, but they do satisfy the canonical
position-momentum commutation relations. Under this transformation, the
Hamiltonian takes the form
\end{subequations}
\begin{align}
\hat{K}_{2}  &  =-\frac{1}{2}\sum_{\vec{k}\neq0}   \sum_{\sigma=1}^2\left(%
\epsilon_{k,\sigma}+\nu_{\sigma}n\lambda_{\vec{k},\sigma\sigma}\right) \nonumber\\
&  \quad\mbox{}+\sum_{\vec{k}\neq0}\sum_{s=\pm}\frac{\Omega_{\vec{k},s}}%
{2}\left(  \hat{P}_{\vec{k},s}\hat{P}_{-\vec{k},s}+\hat{X}_{\vec{k},s}\hat
{X}_{-\vec{k},s}\right)  .
\end{align}

The final transformation, given by%
\begin{equation}
\hat{b}_{\vec{k},s}=\frac{\hat{X}_{-\vec{k},s}+i\hat{P}_{\vec{k},s}}{\sqrt{2}},
\end{equation}
results in the diagonal Hamiltonian,%
\begin{align}
\hat{K}_{2}  &=
\frac{1}{2}\sum_{\vec{k}\neq0}\left(
\sum_{s=\pm}\Omega_{\vec{k},s} - \sum_{\sigma=1}^2 \left(\epsilon_{k,\sigma}+\nu_{\sigma}n\lambda_{\vec{k},\sigma\sigma}\right)
\right)
\nonumber\\
& \quad\mbox{}+\sum_{\vec{k}\neq0}\sum_{s=\pm}\Omega_{\vec{k},s}\hat{b}_{\vec{k},s}^{\dagger}\hat{b}_{\vec{k},s},
\label{eqn:DiagonalizedHamiltonian}
\end{align}
where the operators $\hat{b}_{\vec{k},s}$ satisfy canonical bosonic commutation relations. The first term in Eq.~(\ref{eqn:DiagonalizedHamiltonian}) is a state-independent offset, due to quantum fluctuations, which can be absorbed into the ground state energy. From the second term, it is apparent that $\Omega_{\vec{k},\pm}$ are the two branches of the dispersion relation for this system. Furthermore, in the case where the dressed coupling constant $\Lambda_{\vec{k},12}$ uniformly vanishes, these two dispersion relations reduce to $\omega_{\vec{k},\sigma}$. We will see in Sec.~\ref{sec:ManyBodyProperties} that these two decoupled modes can be interpreted as spin-wave and density-wave modes. The presence of interactions between molecules in different dipole states couples these two modes, and this leads to a rich quasi-particle spectrum in which either density- or spin-wave behavior can dominate.

Since we have explicit expressions for the operator transformations, we can
write the original plane-wave operators $\hat{a}_{\vec{k},\kappa}$ in terms of
the quasi-particle operators $\hat{b}_{\vec{k},s}$ as%
\begin{equation}
\hat{a}_{\vec{k},\kappa}=\sum_{s=\pm}\left(  u_{\vec{k},\kappa s+}\hat
{b}_{\vec{k},s}+v_{-\vec{k},\kappa s}^{\ast}\hat{b}_{-\vec{k},s}^{\dagger
}\right)  , \label{eqn:PlaneWaveToQuasiparticle}%
\end{equation}
where the $u$'s and $v$'s are known as Bogoliubov amplitudes. Many quantities
that characterize the system---such as structure factors and correlation
functions---can be written in terms of these amplitudes, and we therefore
quote the results here. They are
\begin{subequations}
\label{eqn:BogoliubovAmplitudes}%
\begin{align}
u_{\vec{k},1+}  &  =\frac{\Gamma_{\vec{k},1+}^{2}+1}{2\Gamma_{\vec{k},1+}}%
\cos\gamma_{\vec{k}}\cos\eta\nonumber\\
&  \quad\mbox{}-\frac{\Gamma_{\vec{k},2+}^{2}+1}{2\Gamma_{\vec{k},2+}%
}e^{-i\phi_{\vec{k}}}\sin\gamma_{\vec{k}}\sin\eta,\\
u_{\vec{k},1-}  &  =-\frac{\Gamma_{\vec{k},1-}^{2}+1}{2\Gamma_{\vec{k},1-}%
}e^{i\phi_{\vec{k}}}\sin\gamma_{\vec{k}}\cos\eta\nonumber\\
&  \quad\mbox{}-\frac{\Gamma_{\vec{k},2-}^{2}+1}{2\Gamma_{\vec{k},2-}}%
\cos\gamma_{\vec{k}}\sin\eta,\\
v_{-\vec{k},1+}^{\ast}  &  =\frac{\Gamma_{\vec{k},1+}^{2}-1}{2\Gamma_{\vec
{k},1+}}\cos\gamma_{\vec{k}}\cos\eta\nonumber\\
&  \quad\mbox{}-\frac{\Gamma_{\vec{k},2+}^{2}-1}{2\Gamma_{\vec{k},2+}%
}e^{-i\phi_{\vec{k}}}\sin\gamma_{\vec{k}}\sin\eta,\\
v_{-\vec{k},1-}^{\ast}  &  =-\frac{\Gamma_{\vec{k},1-}^{2}-1}{2\Gamma_{\vec
{k},1-}}e^{i\phi_{\vec{k}}}\sin\gamma_{\vec{k}}\cos\eta\nonumber\\
&  \quad\mbox{}-\frac{\Gamma_{\vec{k},2-}^{2}-1}{2\Gamma_{\vec{k},2-}}%
\cos\gamma_{\vec{k}}\sin\eta,
\end{align}
and%
\begin{align}
\label{eqn:BogoliubovAmplitudes}u_{\vec{k},2+}  &  =\frac{\Gamma_{\vec{k}%
,1+}^{2}+1}{2\Gamma_{\vec{k},1+}}\cos\gamma_{\vec{k}}\sin\eta\nonumber\\
&  \quad\mbox{}+\frac{\Gamma_{\vec{k},2+}^{2}+1}{2\Gamma_{\vec{k},2+}%
}e^{-i\phi_{\vec{k}}}\sin\gamma_{\vec{k}}\cos\eta,\\
u_{\vec{k},2-}  &  =-\frac{\Gamma_{\vec{k},1-}^{2}+1}{2\Gamma_{\vec{k},1-}%
}e^{i\phi_{\vec{k}}}\sin\gamma_{\vec{k}}\sin\eta\nonumber\\
&  \quad\mbox{}+\frac{\Gamma_{\vec{k},2-}^{2}+1}{2\Gamma_{\vec{k},2-}}%
\cos\gamma_{\vec{k}}\cos\eta,\\
v_{-\vec{k},2+}^{\ast}  &  =\frac{\Gamma_{\vec{k},1+}^{2}-1}{2\Gamma_{\vec
{k},1+}}\cos\gamma_{\vec{k}}\sin\eta\nonumber\\
&  \quad\mbox{}+\frac{\Gamma_{\vec{k},2+}^{2}-1}{2\Gamma_{\vec{k},2+}%
}e^{-i\phi_{\vec{k}}}\sin\gamma_{\vec{k}}\cos\eta,\\
v_{-\vec{k},2-}^{\ast}  &  =-\frac{\Gamma_{\vec{k},1-}^{2}-1}{2\Gamma_{\vec
{k},1-}}e^{i\phi_{\vec{k}}}\sin\gamma_{\vec{k}}\sin\eta\nonumber\\
&  \quad\mbox{}+\frac{\Gamma_{\vec{k},2-}^{2}-1}{2\Gamma_{\vec{k},2-}}%
\cos\gamma_{\vec{k}}\cos\eta,
\end{align}
where%
\end{subequations}
\begin{equation}
\Gamma_{\vec{k},\sigma s}=\sqrt{\frac{\epsilon_{k,\sigma}}{\Omega_{\vec{k},s}%
}}.
\end{equation}
The Bogoliubov amplitudes satisfy the relations,
\begin{subequations}
\begin{align}
v_{-\vec{k},\sigma s}^{\ast}  &= v_{\vec{k},\sigma s},\\
u_{-\vec{k},\sigma s}^{\ast}  &= u_{\vec{k},\sigma s}.
\end{align}
\end{subequations}
and the orthonormality conditions,%
\begin{subequations}
\begin{align}
\delta_{\kappa\kappa^{\prime}}  &  =\sum_{\sigma=\pm}\left(  u_{\kappa\sigma
}u_{\kappa^{\prime}\sigma}^{\ast}-v_{\vec{k},\kappa\sigma}v_{\vec{k}%
,\kappa^{\prime}\sigma}^{\ast}\right)  ,\\
0  &  =\sum_{\sigma=\pm}\left(  v_{\vec{k},\kappa\sigma}u_{\vec{k}%
,\kappa^{\prime}\sigma}^{\ast}-u_{\vec{k},\kappa\sigma}v_{\vec{k}%
,\kappa^{\prime}\sigma}^{\ast}\right)  .
\end{align}
\end{subequations}

\subsection{Many-body characterization of the ground state and low-energy
excitations}

Access to analytic expressions for the Bogoliubov amplitudes allows us to find
analytic expressions for many important quantities that characterize the
nature of both the ground state and the low energy excitations. The momentum
distribution $\left\langle \hat{n}_{\kappa}\left(  \vec{k}\right)
\right\rangle $, where%
\begin{align}
\hat{n}_{\kappa}\left(  \vec{k}\right)   &  =\hat{\psi}_{\kappa}^{\dagger
}\left(  \vec{k}\right)  \hat{\psi}_{\kappa}\left(  \vec{k}\right) \nonumber\\
&  =\frac{1}{A}\int d^{2}\rho\frac{e^{i\vec{k}\cdot{\boldsymbol\rho}}}%
{\sqrt{A}}\int d^{2}\rho^{\prime}\frac{e^{-i\vec{k}\cdot{\boldsymbol\rho
}^{\prime}}}{\sqrt{A}}\hat{\psi}_{\kappa}^{\dagger}\left(  {\boldsymbol\rho
}\right)  \hat{\psi}_{\kappa}\left(  {\boldsymbol\rho}^{\prime}\right)  ,
\end{align}
can be written as the sum of the condensate density $n_{\kappa}\delta_{\vec
{k},0}$ and the density of fluctuations $\delta\bar{n}_{\vec{k},\kappa}$,
given by%
\begin{equation}
\delta\bar{n}_{\vec{k},\kappa}=\frac{1}{A}\sum_{s=\pm}v_{\vec{k},\kappa
s}^{\ast}v_{\vec{k},\kappa s},
\end{equation}
where we have used the expansions in Eqs. (\ref{eqn:PlaneWaveExpansion}) and
(\ref{eqn:PlaneWaveToQuasiparticle}). The expectation value is taken in the
mean-field ground state, i.e.~the quasi-particle vacuum $\left\vert
0\right\rangle $, where $\hat{b}_{\vec{\kappa},\sigma}\left\vert 0\right\rangle =0$. The total momentum-space density is then%
\begin{equation}
\bar{n}_{\vec{k}}=\sum_{\kappa}\left(  n_{\kappa}\delta_{\vec{k},0}+\delta
\bar{n}_{\vec{k},\kappa}\right)
\end{equation}

Bogoliubov theory requires that the ``leakage'' from the condensate to the non-condensate
(excitation) component be small. The size of this leakage is given by the
quantum depletion, which is the difference between the total density and the
density of the condensate,
\begin{equation}
\delta\bar{n}=\sum_{\kappa}\delta\bar{n}_{\kappa}=\sum_{\kappa}\left(  \bar
{n}_{\kappa}-n_{\kappa}\right)  ,
\end{equation}
where $\bar{n}_{\kappa}$, given by
\begin{equation}
\bar{n}_{\kappa}=\langle\hat{n}_{\kappa}\left(  {\boldsymbol\rho}\right)
\rangle=\langle\hat{\psi}_{\kappa}^{\dagger}\left(  {\boldsymbol\rho}\right)
\hat{\psi}_{\kappa}\left(  {\boldsymbol\rho}\right)  \rangle,
\end{equation}
is the total density of the condensate and fluctuations of component $\kappa$,
and we are working in oscillator units (i.e.~$n_{\kappa}=l^{2}N_{k}/A$). Due
to the translational symmetry in the plane perpendicular to the trap axis,
this quantity is independent of ${\boldsymbol\rho}$ and is hence equal to the
average density $\bar{n}=\sum_{\vec{k}}\bar{n}_{\vec{k}}$. Therefore, the component depletions are given by
\begin{equation}
\delta\bar{n}_{\kappa}=\int\frac{d^{2}k}{\left(  2\pi\right)  ^{2}}\sum
_{s=\pm}v_{\vec{k},\kappa s}^{\ast}v_{\vec{k},\kappa s},
\end{equation}
in the thermodynamic limit. The condition for the Bogoliubov-de Gennes approximation to be valid is that $\delta\bar{n}\ll n$.

To characterize the ground state and low-energy excitations, we are interested in the behavior of the spontaneous fluctuations of the system and of the response of the system to small perturbations. In particular, we will be interested in density fluctuations and spin fluctuations, characterized by the position-space, density-density and spin-spin correlation functions, respectively. Alternatively, we are interested in the response of the system to fluctuations in the external potential and in the applied field, characterized by the corresponding static susceptibilities.

The density-density correlation function $G_{n}^{\left(  2\right)  }\left(
{\boldsymbol\rho}\right)  $ can be written as%
\begin{align}
G_{n}^{\left(  2\right)  }\left(  {\boldsymbol\rho}\right)   &  =\left\langle
\hat{n}\left(  {\boldsymbol\rho}\right)  \hat{n}\left(  \vec{0}\right)
\right\rangle -\left\langle \hat{n}\left(  {\boldsymbol\rho}\right)
\right\rangle \left\langle \hat{n}\left(  \vec{0}\right)  \right\rangle
\nonumber\\
&  \quad\mbox{}-\delta\left(  {\boldsymbol\rho}\right)  \left\langle \hat
{n}\left(  \vec{0}\right)  \right\rangle ,
\end{align}
where $\hat{n}\left(  {\boldsymbol\rho}\right)  $, given by $\hat{n}\left(  {\boldsymbol\rho}\right)  =\sum_{\kappa}\hat{n}_{\kappa
}\left(  {\boldsymbol\rho}\right)$ is the total density of the system at position ${\boldsymbol\rho}$. The term
proportional to the delta function has been included to remove the singular
part of $\left\langle \hat{n}\left(  {\boldsymbol\rho}\right)  \hat{n}\left(
{\vec{0}}\right)  \right\rangle $ at ${\boldsymbol\rho=\vec{0}}$. Roughly
speaking, the function $G_{n}^{\left(  2\right)  }\left(  {\boldsymbol\rho
}\right)  $ gives the relative probability for detecting a particle at
position ${\boldsymbol\rho}$ given that we have already detected a particle at
position $\vec{0}$. Since the system is translationally invariant in the
longitudinal direction, the origin $\vec{0}$ is arbitrary. The spin-spin
correlation function $G_{\Delta}^{\left(  2\right)  }\left(  {\boldsymbol\rho
}\right)  $ can be written as%
\begin{align*}
G_{\Delta}^{\left(  2\right)  }\left(  {\boldsymbol\rho}\right)   &
=\langle\hat{\Delta}\left(  {\boldsymbol\rho}\right)  \hat{\Delta}\left(
\vec{0}\right)  \rangle-\langle\hat{\Delta}\left(  {\boldsymbol\rho}\right)
\rangle\langle\hat{\Delta}\left(  \vec{0}\right)  \rangle\\
&  \quad\mbox{}-\delta\left(  {\boldsymbol\rho}\right)  \sum_{\kappa}%
d_{\kappa}^{2}\left\langle \hat{n}_{\kappa}\left(  \vec{0}\right)
\right\rangle ,
\end{align*}
where $\hat{\Delta}\left(  {\boldsymbol\rho}\right)  $, given by%
\begin{equation}
\hat{\Delta}\left(  {\boldsymbol\rho}\right)  =\sum_{\kappa}d_{\kappa}\hat
{n}_{\kappa}\left(  {\boldsymbol\rho}\right)  ,
\end{equation}
is the polarization operator. Again, the term proportional to the delta
function has been included to remove the singular part of $\langle\hat{\Delta
}\left(  {\boldsymbol\rho}\right)  \hat{\Delta}\left(  {\boldsymbol0}\right)
\rangle$ at ${\boldsymbol\rho=\vec{0}}$. We have subtracted off the
long-distance behaviors, given by%
\begin{equation}
\left\langle \hat{n}\left(  {\boldsymbol\rho}\right)  \right\rangle
\left\langle \hat{n}\left(  \vec{0}\right)  \right\rangle =\bar{n}^{2},
\end{equation}
and%
\begin{equation}
\langle\hat{\Delta}\left(  {\boldsymbol\rho}\right)  \rangle\langle\hat
{\Delta}\left(  \vec{0}\right)  \rangle=d^{2}\left(  \bar{n}_{\uparrow}%
-\bar{n}_{\downarrow}\right)  ^{2}.
\end{equation}

The two correlation functions can be written in terms of the Bogoliubov
amplitudes defined in Eq.~(\ref{eqn:BogoliubovAmplitudes}) by expanding the
field operators using Eqs. (\ref{eqn:PlaneWaveExpansion}) and
(\ref{eqn:PlaneWaveToQuasiparticle}), and the results are%
\begin{align}
G_{n}^{\left(  2\right)  }\left(  {\boldsymbol\rho}\right)   &  =n\int
\frac{d^{2}k}{\left(  2\pi\right)  ^{2}}e^{i\vec{k}\cdot{\boldsymbol\rho}%
}\left(  S_{n}\left(  \vec{k}\right)  -1\right)  ,\\
G_{\Delta}^{\left(  2\right)  }\left(  {\boldsymbol\rho}\right)   &
=nd^{2}\int\frac{d^{2}k}{\left(  2\pi\right)  ^{2}}e^{i\vec{k}\cdot
{\boldsymbol\rho}}\left(  S_{\Delta}\left(  \vec{k}\right)  -1\right)  ,
\end{align}
where we have defined the density and polarization structure factors,%
\begin{align}
S_{n}\left(  \vec{k}\right)   &  =\sum_{\kappa,\kappa^{\prime}}S_{\kappa
\kappa^{\prime}}\left(  \vec{k}\right)  ,\\
S_{\Delta}\left(  \vec{k}\right)   &  =\sum_{\kappa,\kappa^{\prime}}\left(
-1\right)  ^{1-\delta_{\kappa\kappa^{\prime}}}S_{\kappa\kappa^{\prime}}\left(
\vec{k}\right)  ,
\end{align}
and%
\begin{align}
S_{\kappa\kappa^{\prime}}\left(  \vec{k}\right)   &  =\left(  -1\right)
^{1-\delta_{\kappa\kappa^{\prime}}}\sqrt{\nu_{\kappa}\nu_{\kappa^{\prime}}%
}\sum_{s=\pm}\nonumber\\
&  \quad\mbox{}\times\left(  u_{\kappa s}\left(  \vec{k}\right)  +v_{\kappa
s}\left(  \vec{k}\right)  \right)  \left(  u_{\kappa^{\prime}s}^{\ast}\left(
\vec{k}\right)  +v_{\kappa^{\prime}s}^{\ast}\left(  \vec{k}\right)  \right)  .
\end{align}

Up to constants, these structure factors are merely the Fourier transforms of
the two-point correlation functions. They act as the connection between the
correlation functions that characterize fluctuations of the ground state and
the static susceptibility functions that characterize the response of the
system to external perturbations. See Appendix \ref{app:LinearResponseTheory} for careful definitions of the response functions. In
Sec.~\ref{sec:ManyBodyProperties}, we use these expressions to clarify the
nature of the instabilities that emerge at large densities for both high
and low fields.

\section{Properties of the Mean-Field Ground State\label{sec:GroundStateProperties}}

In the remaining sections of this paper, we apply the general theory outlined above to the specific system of a BEC of dipolar molecules interacting via dipole-dipole interactions, trapped in quasi-2D, and experiencing a constant electric polarizing field oriented perpendicular to the plane. We begin our discussion of the behavior of this system by investigating the properties of the mean field ground state, including the extrema of the variational energy functional and the polarization of the ground state. We employ the Gaussian ansatz for the axial wave functions (Eq.~\ref{eqn:GaussianAnsatz}).

The ground state energy functional (Eq.~\ref{eqn:GroundStateEnergyGaussian}) can be written as
\begin{align}
K_{0}  &  =N\frac{\hbar\omega}{2}-N\hbar\omega\beta\gamma\left(  \nu
_{\uparrow}-\nu_{\downarrow}\right)  +N\hbar\omega\gamma\cos\theta\sqrt
{\nu_{\uparrow}\nu_{\downarrow}}\nonumber\\
&  \quad\mbox{}+2\sqrt{2\pi}N\hbar\omega D \left(  \nu_{\uparrow}%
-\nu_{\downarrow}\right)  ^{2} \label{eqn:GroundStateEnergySimple}%
\end{align}
where $\gamma={2h_{c}}/{\hbar\omega}$ is the effective zero-field splitting of the molecule, $\beta={h_{0}}/{2h_{c}}$ is the electric field coupling---i.e.~half the linear Stark shift---relative
to the zero-field splitting, and $D={nd^{2}}/{3\hbar\omega l^{3}}$ is an effective density-dependent
interaction parameter. In the 3D homogeneous case, a gas of dipolar molecules
is automatically unstable but can be stabilized via a repulsive contact
interaction characterized by a scattering length $a$. The condition for
stability is exactly $D<na/l$~\cite{Santos2002}. In the quasi-2D case,
however, the gas is stabilized by the presence of the trapping potential
rather than a repulsive contact interaction.

\begin{figure}[tb]
\includegraphics[width=86mm]{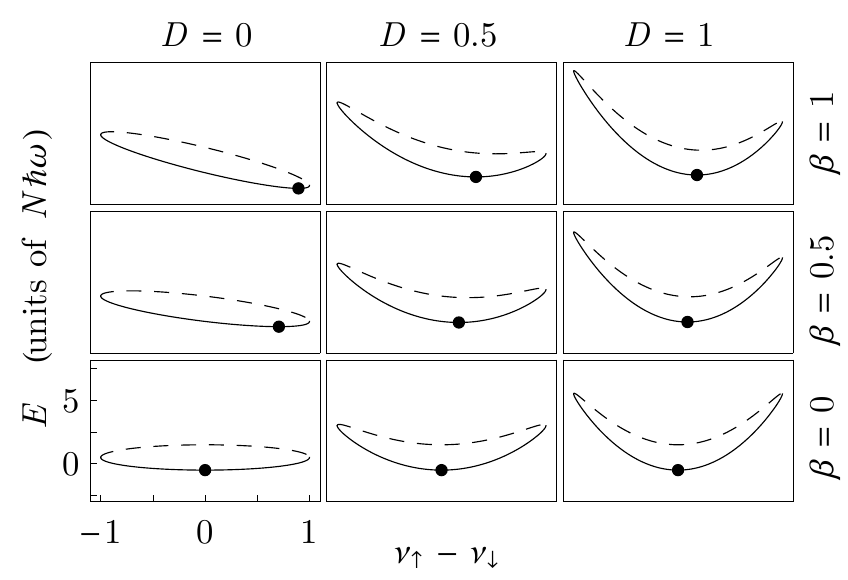}
\caption{Ground state energy functional as a function of the normalized polarization
$\nu_{\uparrow}-\nu_{\downarrow}$ for $\theta=\pi$ (solid), $\theta=0$ (dashed), $\beta=0$, $0.5$, and $1$, and $D=0$, $1$, and $2$. The system at all times has a single ground state (indicated as a point in the plot) that is driven towards large polarization by the external field and zero polarization by the inter-particle interactions.}
\label{fig:EnergyFunctional}
\end{figure}

The energy functional exhibits markedly different behaviors in different
parameter regimes. This is illustrated in Fig.~\ref{fig:EnergyFunctional},
where the two branches of the energy functional are plotted as a function of
the polarization $\nu_{\uparrow}-\nu_{\downarrow}$ for $\gamma=2$, with the $\theta=\pi$
branch as a solid curve and the $\theta=0$ branch as a dashed curve. In all
parameter regimes, there is a single global minimum in the $\pi$ branch,
indicating the existence of a universal ground state. This is in contrast to
Fig.~1 in Ref.~\cite{Tommasini03} where there can exist two minima and one
maximum in the $\pi$-branch and is due to the fact that the interactions
always drive the system towards zero polarization. It is only the external
field that can increase the population imbalance, as shown in the subplots of Fig.~\ref{fig:EnergyFunctional} in which there are no interactions ($D=0$).
As the external field is increased, the ground state becomes more polarized in
the direction of the external field. The gas is essentially fully polarized by
the time $\beta\approx2$. In addition, the energy has a global maximum in the
$0$ branch. This state is highly polarized anti-parallel to the external field
and is thus significantly higher in energy than the universal ground state.

The situation changes when the interaction strength---or equivalently, the
density---is increased. The interaction drives the mean-field ground state in
the $\pi$-branch toward zero polarization due to attractive in-plane interactions
between unlike dipoles and repulsive interactions between like dipoles. In
addition, at a particular interaction strength, the global maximum in the
$0$-branch splits into two local maxima that bound one local minimum, which is
actually a saddle point due to this being the branch that maximizes the energy
in $\theta$. This pushes the global maximum to even larger negative
polarization, while the $0$-branch minimum remains of slightly larger
polarization than the global minimum for all values of $D$ larger than this threshold.

\begin{figure}[tb]
\includegraphics[width=86mm]{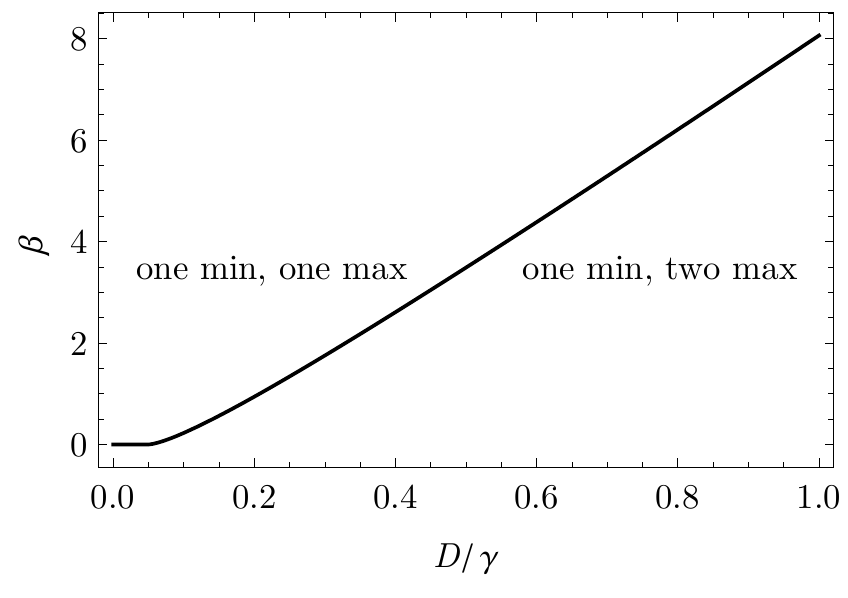}
\caption{Critical threshold for the existence of three local extrema in the $\theta=0$ branch of the ground state energy functional.}
\label{fig:CriticalPoints}
\end{figure}

This threshold value of $D$ is actually a function of $\beta$, and the
relationship can be computed by minimizing both the the energy and the first
derivative of the energy, resulting in the relationship
\begin{align}
\beta &  =\frac{1}{2}\left(  4\sqrt[3]{2\pi}\left(  \frac{D}{\gamma}\right)
^{2/3}-1\right)  ^{3/2}~.
\end{align}
This relationship is plotted in Fig.~\ref{fig:CriticalPoints}.
For small values of the effective interaction strength $D$, there is only ever
one extremum on the upper branch, but for values above a critical value
$D_{c} = \gamma/16\pi$, there are two possibilities for the number of extrema, depending on the value of $\beta$. These extra extrema in the ground-state energy functional may be of additional physical importance; for instance, it is possible that such states correspond to dynamically stable states along the lines of Ref.~\cite{Bernier2014a}. This is left for future study.

\begin{figure}[tb]
\includegraphics[width=86mm]{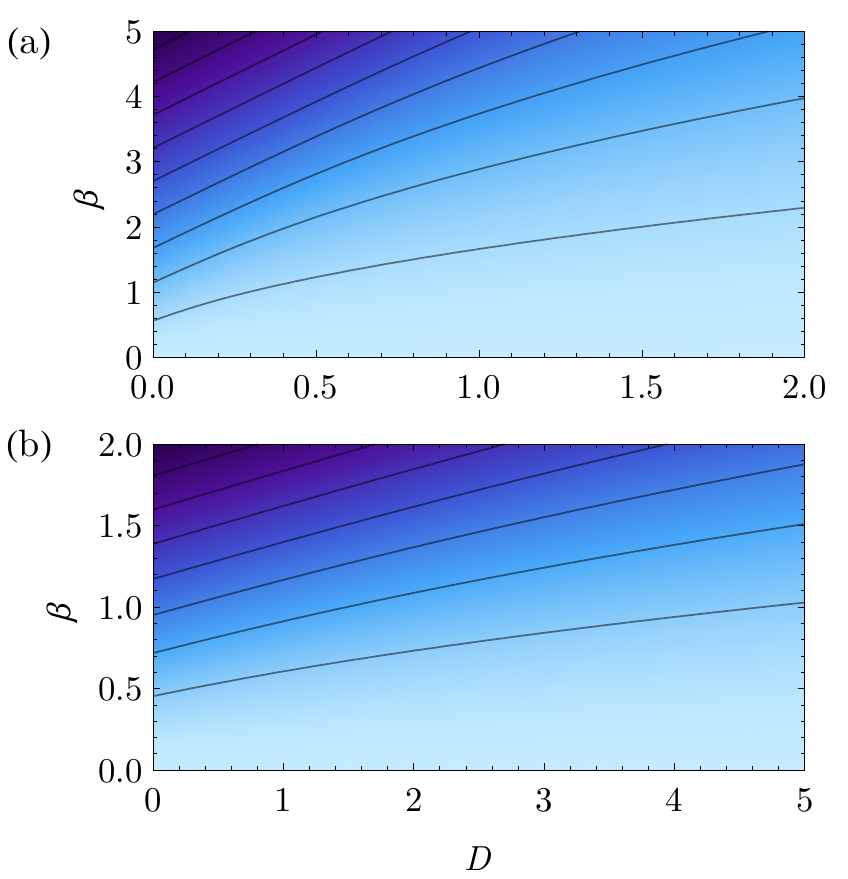}
\caption{(Color online.) Contour plot of the ground state energy as a function of $\beta$ and $D$ for (a) $\gamma=2$ and (b) $\gamma=20$. Darker shades indicate lower energy. (a) The contours march from $-4\hbar\omega$ to 0 in steps of $0.5\hbar\omega$ from darker to lighter shades. (b) The contours march from $-18\hbar\omega$ to $-6\hbar\omega$ in steps of $2\hbar\omega$ from darker to lighter shades.}
\label{fig:GroundStateEnergy}
\end{figure}

In Fig.~\ref{fig:GroundStateEnergy}, we have plotted the energy of the global
minimum---that is, the ground state energy---as a function of $\beta$ and $D$
for $\gamma=2$ and $\gamma=20$. It is apparent from these plots that the energy decreases with field strength, indicating that the dipoles are aligning with the external field and displaying a Stark shift, and that the energy increases with interaction strength, indicating that the interactions are driving the system towards a state of zero polarization.

\begin{figure}[tb]
\includegraphics[width=86mm]{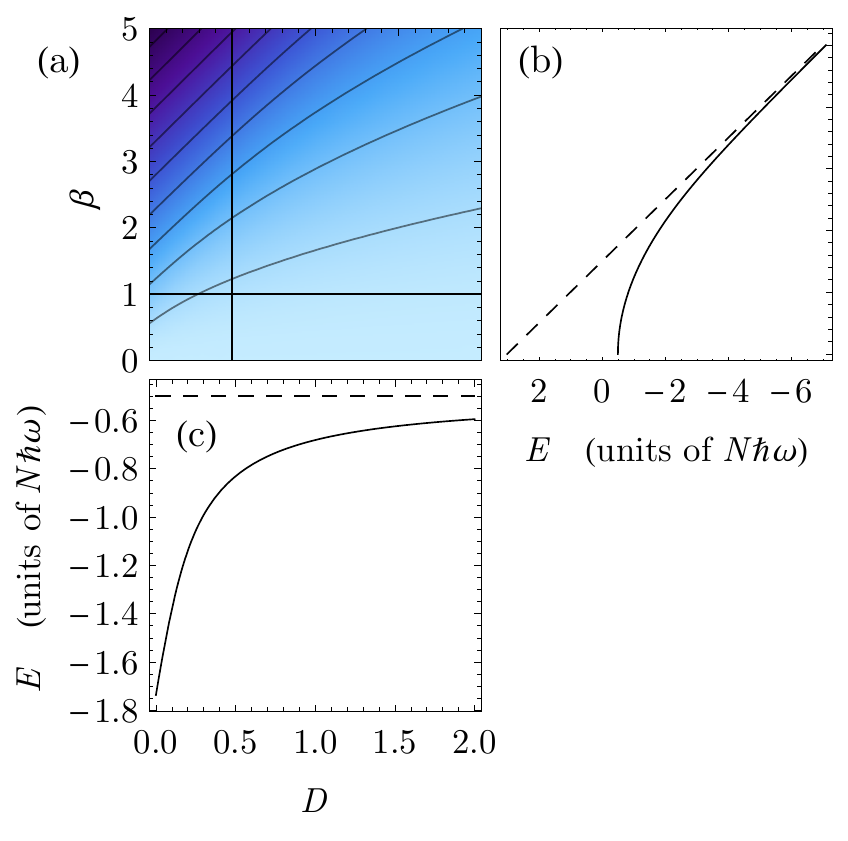}
\caption{(Color online.) (a) Contour plot of the ground state energy for $\gamma=2$ as a function of $\beta$ and $D$. (b) The ground state energy is plotted as a function of $\beta$ for $D=0.5$ (vertical line in (a)). In the asymptotic limit $\beta\to\infty$, the energy decreases linearly with $\beta$ (dashed line), which is the well-known linear Stark shift. (c) The ground state energy is plotted as a function of $D$ for $\beta=1$ (horizontal line in (a)). The energy approaches a universal (at fixed $\gamma$) limit as $D\to\infty$ (dashed line), indicating that interactions have driven the system into a state of zero polarization where both the effects of the external field and of interactions vanish as a result of screening.}
\label{fig:GroundStateEnergyLimits}
\end{figure}

These two limits are clarified in Fig.~\ref{fig:GroundStateEnergyLimits}, in which we plot the ground state energy as a function of $D$ for $\beta = 1$ and $\beta$ for $D=0.5$. We show the limiting cases as dashed lines. In (b), we can see that in the limit of large $\beta$, the ground state energy decreases linearly with $\beta$. In the asymptotic limit where $\beta$ is large, the gas is fully polarized, and the ground state energy functional (Eq.~\ref{eqn:GroundStateEnergySimple}) becomes
\begin{equation}
\frac{K_{0}}{N\hbar\omega}\rightarrow
\frac{1}{2}+2\sqrt{2\pi}D-\beta\gamma,
\end{equation}
which is exactly the equation for the dashed line. The energy decreases linearly with the external field, which is the well-known linear Stark shift.

On the other hand, in (c), it appears that the energy converges to a constant value in the limit of large $D$. Indeed, in this limit, the energy is given exactly by
\begin{equation}
\lim_{D\rightarrow\infty}\frac{K_{0}}{N}=\hbar\omega\left(  \frac{1}{2}%
-\frac{\gamma}{2}\right) ,
\end{equation}
which is the equation for the dashed line. Tellingly, this quantity is independent of $\beta$. This energy is exactly the energy of a single particle in the ground state of the trap plus the zero-field ground state energy of the molecule. The strong interactions drive the system to a state of zero polarization, and in this case the gas is perfectly screened, eliminating the effects of both the external field and the interactions.

\begin{figure}[tb]
\includegraphics[width=86mm]{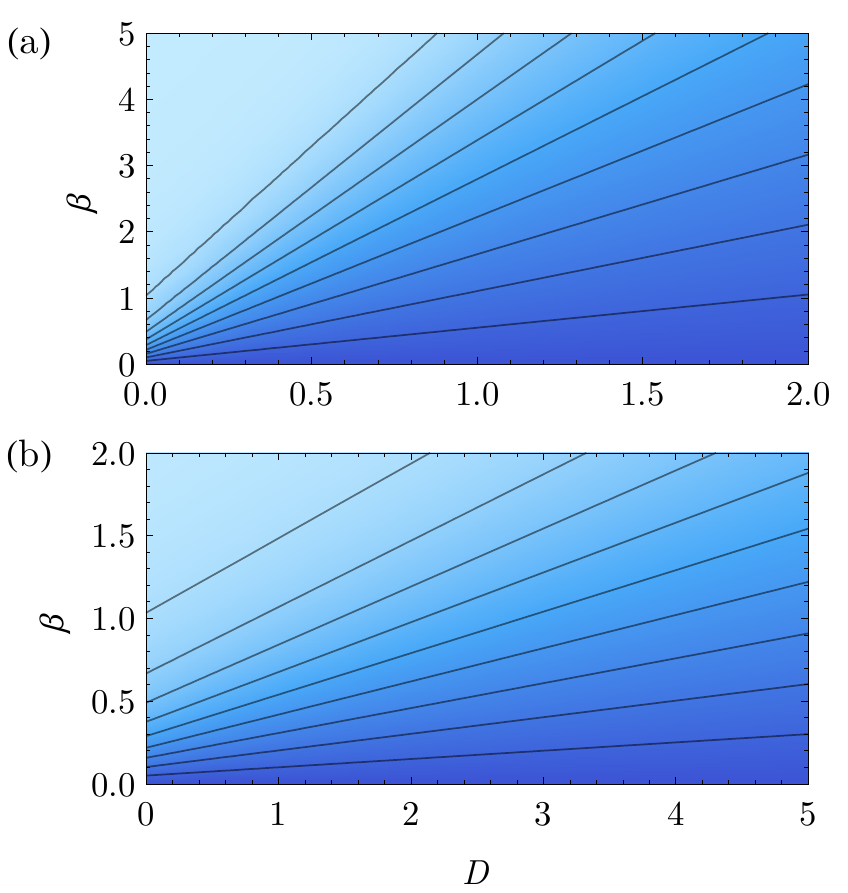}
\caption{(Color online.)
Contour plot of the ground-state polarization ($\nu_{\uparrow}-\nu_{\downarrow}$) as a function of $\beta$ and $D$ for (a)
$\gamma=2$ and (b) $\gamma=20$. The contours march from
$\nu_{\uparrow}-\nu_{\downarrow}=0$ along the $\beta=0$ axis
to $\nu_{\uparrow}-\nu_{\downarrow}=0.9$ in steps of 0.1.
The ground state is driven toward zero polarization (dark shades) with increasing $D$
and toward $\nu_{\uparrow}-\nu_{\downarrow}=1$ (light shades) with increasing $\beta$.}
\label{fig:GroundStatePolarization}
\end{figure}

It is clear from the preceding that there is a competition between the
external field, which drives the system towards a larger polarization, and the
interactions, which drive the system towards zero polarization. This can be
seen explicitly in Fig.~\ref{fig:GroundStatePolarization}, in which we have
plotted the polarization as a function of $\beta$ and $D$ for $\gamma =2$ and $\gamma=20$. It is clear immediately that the interaction reduces the polarization whereas the external
field increases it. Furthermore, the polarization increases  with
$\beta$ until it saturates at the maximum value of $Nd$ where the gas is fully
polarized, and the polarization is zero for large enough $D$. This behavior is
exactly the behavior of a dielectric material in which the internal fields
created by the individual molecules partially cancel out the applied field,
self-consistently leading to smaller dipole moments. In the semi-classical
picture, it is the fields generated by the molecules that give rise to the
dielectric behavior. Here, the dielectric behavior manifests as a
competition of the two energy scales associated with the inter-particle
interactions and the interaction between a single molecule and the external field.

At the mean-field level, the gas is perfectly stable in the ground state, and
increasing the interaction strength by increasing the density merely drives
the system into a perfectly screened state in which the polarization is zero.
However, it is well known that in the case of a fully polarized, dipolar gas
in quasi-2D, an instability known as a density-wave rotonization emerges at a
critical density~\cite{Wilson2008a}. The instability arises due to the fact that
at large densities, the energy cost associated with the harmonic trapping is
not large enough to keep the molecules from locally piling up end-to-end.
This instability occurs in our model in the limit of large fields, $\beta\gg\gamma$,
where the gas is fully polarized. At small $\beta$, however, increasing density
drives the system very quickly toward zero polarization, stabilizing the gas
against the collapse just described. Nonetheless, an instability arises that
has a markedly different character than the one just described~\cite{Wilson14}. This instability is a polarization wave instability in which the macroscopic, mean-field polarization is zero, but mesoscopic spin fluctuations arise when it becomes energetically favorable for nearby domains of spins to be anti-aligned. The investigation of the spontaneous fluctuations that emerge in the system that give rise to these instabilities is the subject of the next section.

\section{Many-Body Characterization of the Ground State and Low-Energy Excitations\label{sec:ManyBodyProperties}}

We have seen that there is a competition between the energy scales associated with the inter-particle dipole-dipole interactions and the molecule-field interaction: the external field drives the system towards maximum polarization whereas the interactions drive the system towards zero polarization. There is an additional competition between the attractive interactions between aligned dipoles lined up end-to-end and the external trapping potential confining the molecules to be in-plane. Finally, there is a competition between attractive interactions between in-plane anti-aligned dipoles and the energy cost of flipping a single dipole against the field. As a result of these competing energy scales, mesoscopic fluctuations in both the density and polarization arise. In this section, we investigate the consequences of this behavior. We first investigate the dispersion relations and identify roton-like features that soften as the interaction strength increases, which lead to the instabilities discussed briefly in the last section. We move on to characterizing these instabilities by way of response functions and correlation functions. 

\begin{figure}[tb]
\includegraphics[width=86mm]{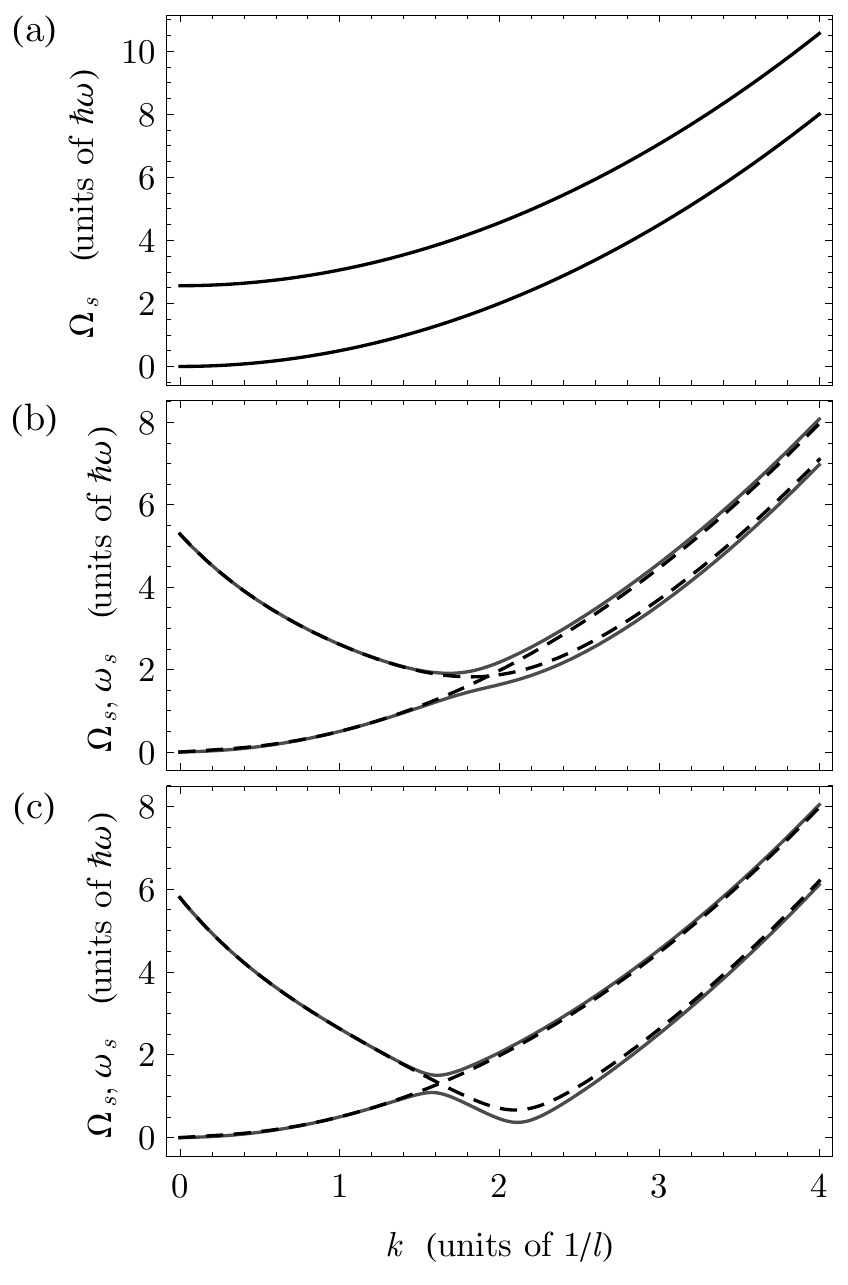}
\caption{Full and decoupled dispersion relations $\Omega_{\vec{k},s}$ (solid) and $\omega_{\vec{k},\sigma}$ (dashed), respectively, plotted for $\gamma=2$ and $\beta=0.4$ at (a) $D=0$, (b) $D=0.6$ and (c) $D=0.74$. (a) For small interaction strengths, the dispersion relations consist essentially of two free-particle branches. (b) Interactions introduce a minimum in the upper branch at finite momentum, and the upper branch will cross the lower branch for large enough $D$, creating an avoided crossing. (c) This behavior results in a roton-like feature that softens for increasing $D$.}
\label{fig:DispersionLowBeta}
\end{figure}

As an issue of nomenclature, we will use the terms spin-wave and polarization wave interchangeably in the following discussion. ``Polarization-wave'' is the correct terminology in the context of this paper, since we are discussing a gas of polar molecules polarized by an external electric field. However, the formalism carries over exactly for a spin-$\frac{1}{2}$ system with long-range interactions, in which the term ``spin-wave'' is relevant.

\subsection{Dispersion relations}

\begin{figure}[tb]
\includegraphics[width=86mm]{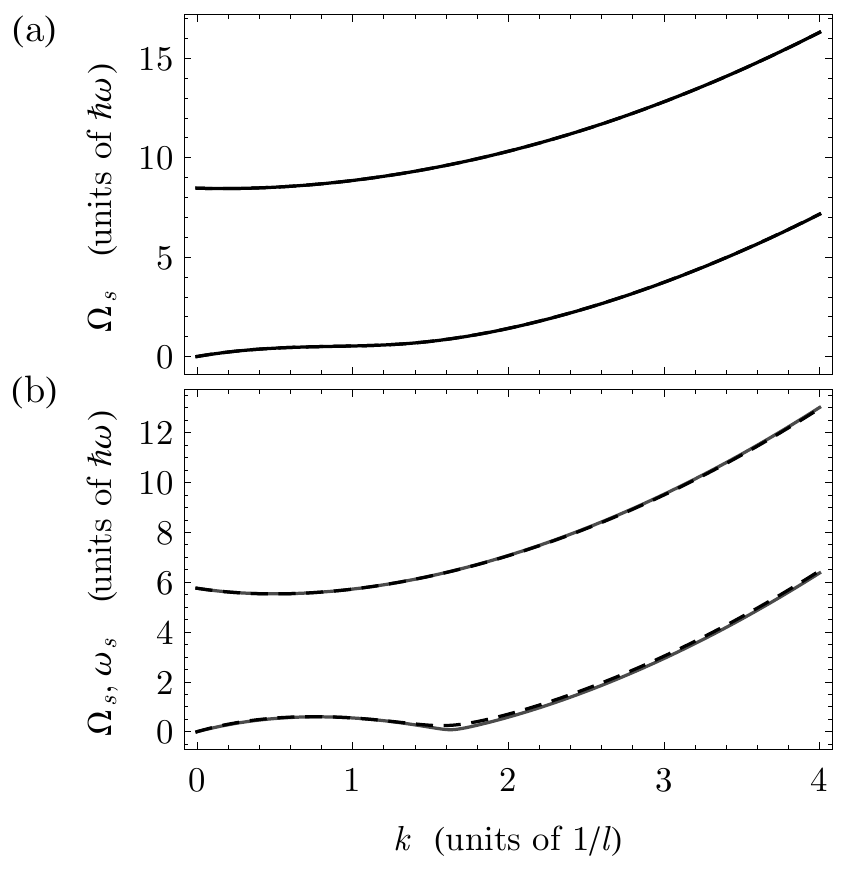}
\caption{Full and decoupled dispersion relations $\Omega_{\vec{k},s}$ (solid) and $\omega_{\vec{k},\sigma}$ (dashed), respectively, plotted for $\gamma=2$ and $\beta=3$ at (a) $D=0.2$ and (b) $D=0.388$. As shown by the almost-perfect overlap between the full and decoupled dispersion curves, the density-wave and spin-wave dispersions are decoupled in this limit, and the lower branch is purely density-wave in character.}
\label{fig:DispersionHighBeta}
\end{figure}

If we carefully examine the diagonalization procedure outlined in Sec.~\ref{sec:ManyBodyTheory}, we can conclude that the two quasiparticle modes characterized by the full dispersion relations $\Omega_{\vec{k},\pm}$ arise from coupling the two modes characterized by $\omega_{\vec{k},\sigma}$ through the dressed interaction parameter $\Lambda_{\vec{k},12}$. In the limit as $\beta\rightarrow0$, the gas has zero polarization and is therefore perfectly screened. It is straight-forward to show that in this limit, the coupling $\Lambda_{\vec{k},12}$ vanishes, and so the dispersion relations are exactly given by $\omega_{\vec{k},\sigma}$. In the limit of no interactions ($D=0$), these are just two free-particle dispersions gapped at zero momentum due to the linear coupling. These modes correspond exactly to the symmetric and anti-symmetric superpositions of $\left\vert \uparrow\right\rangle $ and $\left\vert \downarrow\right\rangle $---that is, they are just the opposite parity eigenstates of the molecule that are coupled by the external field (see Sec.~\ref{sec:SingleMoleculeTheory}). As $D$ increases, the upper branch $\omega_{\vec{k},1}$ develops a minimum at finite momentum. When $\beta\neq0$, the modes $\omega_{\vec{k},\sigma}$ are coupled and above a critical interaction strength undergo an avoided level crossing at some finite momentum. The combination of the avoided crossing and the minimum in the upper branch leads to the emergence of a roton-maxon-like feature in the lower branch of the dispersion. As long as the crossing is narrow, the character of the lower-energy state switches at the avoided crossing; at low momenta, the lower branch is anti-symmetric, whereas at high momenta above the crossing, the lower branch is symmetric. This behavior is shown in Fig.~\ref{fig:DispersionLowBeta}, where we have plotted both $\Omega_{\vec{k},\pm}$ and $\omega_{\vec{k},\sigma}$ at $\gamma=2$ and $\beta=0.4$ for $D=0$, $D=0.6$, and $D=0.74$. At $D\approx0.7459$, the energy of the roton-like feature is zero, and above this value the dispersion relation is complex, signifying a dynamical instability.

\begin{figure}[tb]
\includegraphics[width=86mm]{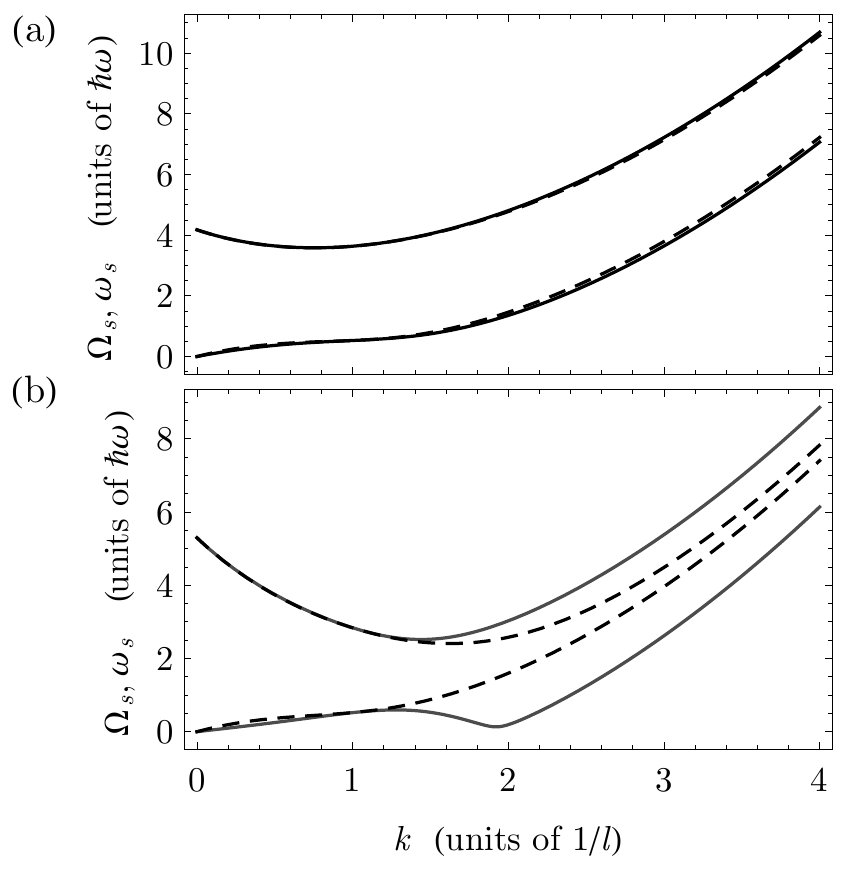}
\caption{Full and decoupled dispersion relations $\Omega_{\vec{k},s}$ (solid) and $\omega_{\vec{k},\sigma}$ (dashed), respectively, plotted for $\gamma=2$ and $\beta=1.75$ at (a) $D=0.3$ and (b) $D=0.645$. The spin wave and density wave dispersion relations $\omega_{\vec{k},\sigma}$ are strongly coupled in this regime, leading to a mixed density and spin wave roton feature.}
\label{fig:DispersionMidBeta}
\end{figure}

In the high-field limit (Fig.~\ref{fig:DispersionHighBeta}), the situation is markedly different. In this case, the gas is fully polarized, and the dispersion relation has two branches that are well-separated in energy. The upper branch is approximately a free-particle dispersion but nonzero at zero momentum. The lower branch, on the other hand, displays a roton-maxon character that is already very well-understood. This is exactly the density-wave roton that arises in a gas of fully polarized molecules, and it is the softening of this roton that leads to the instability discussed previously (see Sec.~\ref{sec:GroundStateProperties} and Ref.~\cite{Bohn09b}). The mechanism by which this roton arises is very different than that of the roton that arises at small fields. In this high-field limit, the two dispersions $\omega_{\vec{k},\sigma}$ are essentially decoupled due to the large gap. Therefore, $\Omega_{\vec{k},\pm}$ and $\omega_{\vec{k},\sigma}$ coincide, as can be seen in Fig.~\ref{fig:DispersionHighBeta}, where we have plotted both at $\gamma=2$ and $\beta=3$ for $D=0.2$, and $D=0.388$. Because the two modes are decoupled, the roton does not arise due to the avoided crossing between the two branches that correspond to spin-wave and density-wave modes, and in fact the lower branch of the dispersion always has density-wave character. As is well known, the roton already appears in the dispersion for the single-component, fully-polarized dipolar BEC.

Finally, in Fig.~\ref{fig:DispersionMidBeta}, we have plotted both $\Omega_{\vec{k},\pm}$ and $\omega_{\vec{k},\sigma}$ at $\gamma=2$ and $\beta=1.75$ for $D=0.3$, and $D=0.645$. These parameters correspond to the cross-over region between the high-field and low-field limits (filled circles in Fig.~\ref{fig:Stability}). As can be seen from Fig.~\ref{fig:DispersionMidBeta}(b), the density-wave and spin-wave modes are appreciably coupled in this regime, leading to a widening of the gap between the upper and lower branches of the dispersion relation. This coupling leads to a roton-like feature in the lower branch that continues to soften as the interaction strength is increased. Near $D\approx0.645$, the dispersion relation goes complex, leading again to a dynamical instability. This instability has both density-wave and spin-wave character due to the appreciable coupling between the two branches.

Based on this discussion, we have identified three distinct mechanisms for the formation of a roton-like feature in the dispersion relation. This suggests that the physical mechanisms for these instabilities are different, and we continue to explore this in the next sections.

\subsection{Stability}

\begin{figure}[tb]
\includegraphics[width=86mm]{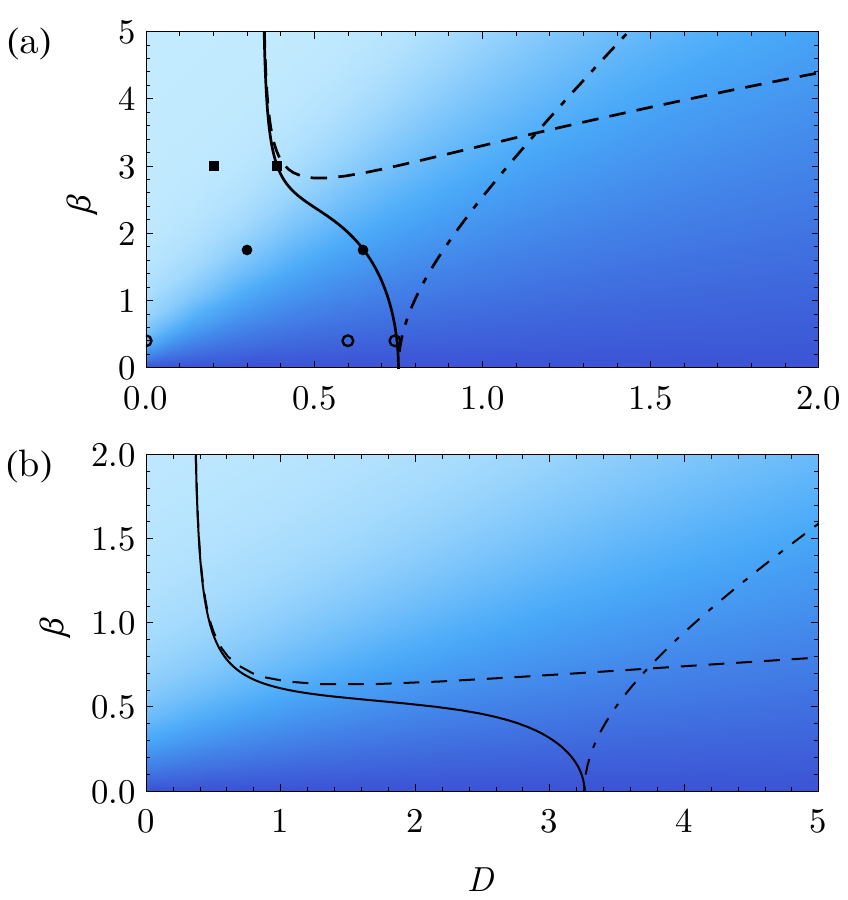}
\caption{(Color online.) The stability diagram for (a) $\gamma=2$ and (b) $\gamma=20$, plotted above the density plots for the ground-state polarization from Fig.~\ref{fig:GroundStatePolarization}. The solid curve is the
stability threshold for the full system, and the dashed and dot-dashed curves
are the stability thresholds corresponding to the density-wave and spin-wave
instabilities, respectively. The points indicated correspond to parameters
chosen in Figs.~\ref{fig:DispersionLowBeta}--\ref{fig:StructureHighField}.
}
\label{fig:Stability}
\end{figure}

In each of the three regimes discussed above, a roton-like feature in the lower branch of the dispersion softens as the interaction strength is increased. At a critical value of $D$, the energy of the roton minimum is zero, and above this threshold, the dispersion is complex, indicating the onset of a dynamical instability. We can map out the stability threshold by finding the values $D$ at which $\Omega_{\vec{k},-}$ goes complex for each $\beta$. In Fig.~\ref{fig:Stability}, we have plotted the stability threshold for the cases $\gamma=2$ and $\gamma=20$. In order to understand the nature of the instabilities, it is necessary to understand the behavior of the decoupled modes $\omega_{\vec{k},1}$ and $\omega_{\vec{k},2}$ that correspond to spin waves and density waves, respectively. In Fig.~\ref{fig:Stability}, we have mapped the values $(D,\beta)$ at which $\omega_{\vec{k},1}$ and $\omega_{\vec{k},2}$ go complex. These are shown as a dashed curve ($\omega_{\vec{k},2}$) and a dot-dashed curve ($\omega_{\vec{k},1}$).

At high fields, the stability threshold matches that corresponding to $\omega_{\vec{k},2}$. This is the limit of the density-wave rotonization. As the field is decreased, the density-wave threshold moves out to higher densities. The molecules become less polarized at lower fields, leading to a smaller effective interaction strength. Therefore, higher densities are necessary in order to access the instability. At small enough fields, no density-wave instability occurs, because $D$ is large enough that the gas has been driven to a state of near-zero polarization, which screens the interaction. This leads to a region of stability for large $D$ and large $\beta$.

At low fields, the stability threshold matches that corresponding to $\omega_{\vec{k},1}$. As we will see, the spontaneous fluctuations of the system near this threshold are purely spin-wave in character, meaning that there is a separation of domains of anti-aligned spins. These domains attract, and at large enough interaction strengths, cause the gas to destabilize. At higher fields, $D$ must be large enough to drive the system into a state of near-zero polarization before this instability sets in. The spin-wave threshold therefore gets pushed out to larger densities at higher fields, which is why this instability disappears in the higher-field regime.

In the cross-over region, the stability threshold smoothly interpolates the high-field and low-field limits, leaving a large unstable region in a parameter regime that is stable in the case where the density- and spin-wave modes are decoupled. Our picture of the cross-over region is then one of a spin-wave-assisted density-wave instability.

\subsection{Momentum distributions, structure factors, and correlation functions}

\begin{figure}[tb]
\includegraphics[width=86mm]{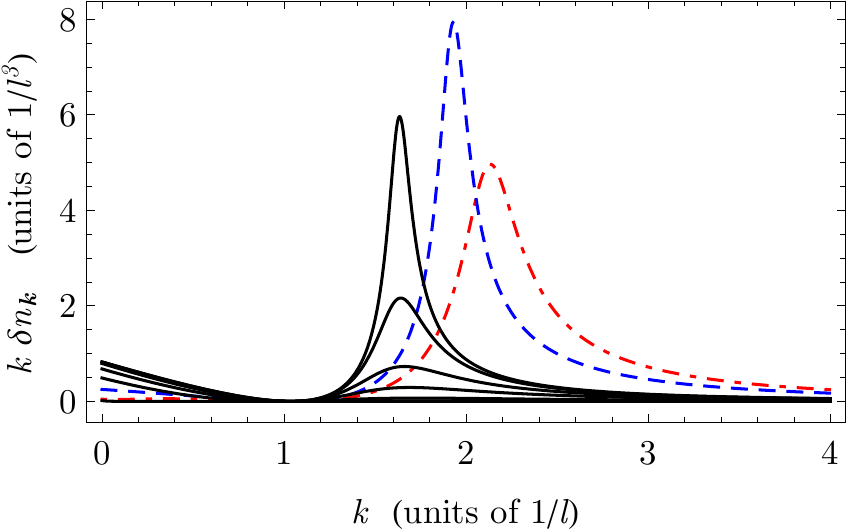}
\caption{(Color online.) Non-condensate momentum distribution (quantum depletion) for $\gamma=2$ and $\beta=3$ with $D=0,~0.1,~0.2,~0.3,~0.35,~0.38,~0.388$ (solid, black); $\beta=1.75$ with $D=0.645$ (dashed, blue); and $\beta=0.4$ with $D=0.74$ (dot-dashed, red). The largest values of $D$ correspond to the points near the stability thresholds in Fig.~\ref{fig:Stability}. The peak in the distribution rises as the interaction strength increases, and the position of the peak moves towards larger $k$ as $\beta$ decreases.}
\label{fig:MomentumDistributions}
\end{figure}

Since the instabilities are the result of spontaneous fluctuations setting up spin- and density-waves of characteristic wavelengths, the instabilities should be signaled by divergences in the non-condensate momentum distribution, i.e.~the depletion. In Fig.~\ref{fig:MomentumDistributions}, we have plotted the momentum distributions $\delta\bar{n}_k$ for a range of values of $D$ from $0$ to near the stability threshold for the high field, mid-field, and low-field cases. We have included a factor of $k$ that comes from the measure $d^{2}k$, which renders $\delta\bar{n}_k$ finite at zero momentum. As $D$ increases, a peak at the position of the roton minimum in the dispersion emerges, and this peak diverges as $D$ goes to the critical value at the stability threshold. As the field strength $\beta$ is increased, the position of the peak moves towards smaller $k$.

\begin{figure}[tb]
\includegraphics[width=86mm]{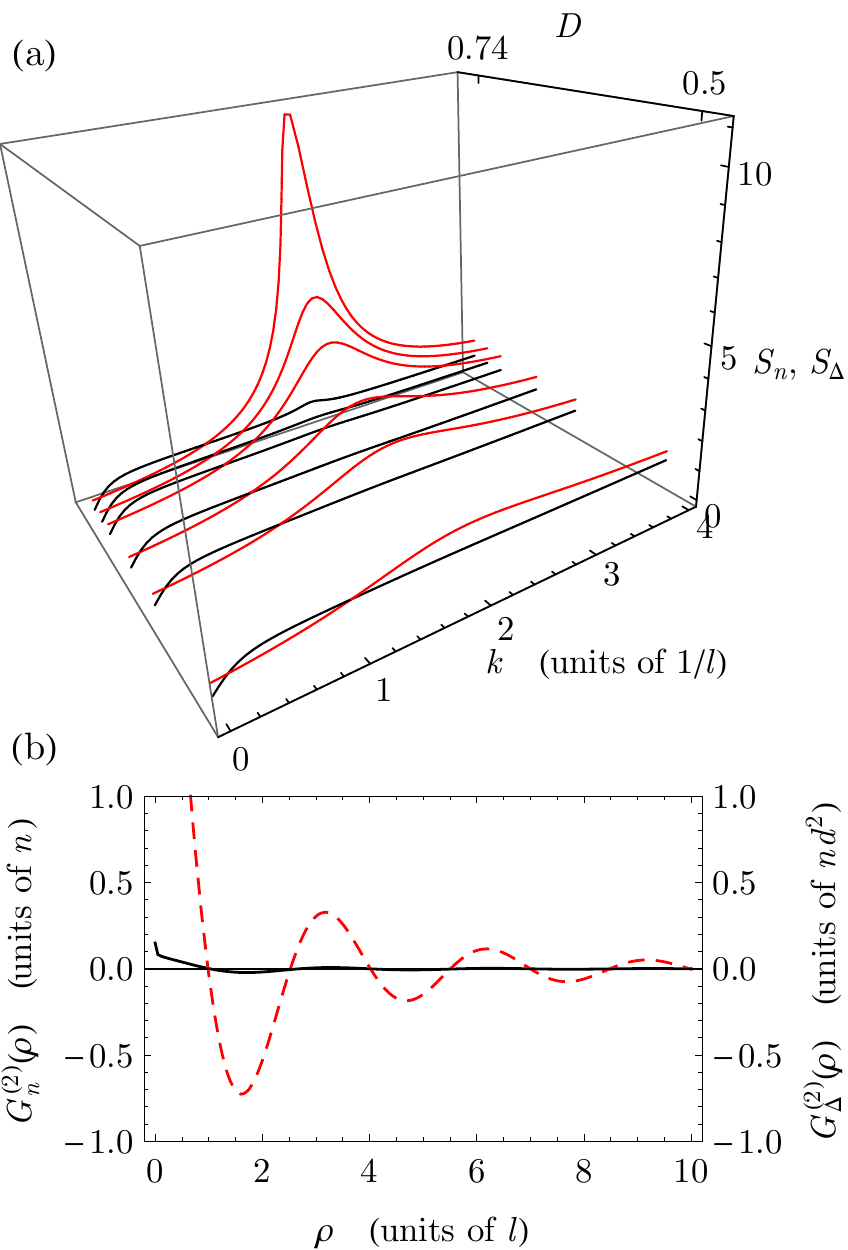}
\caption{(Color online.) (a) Density (black) and spin (red) static structure factors for $\gamma=2$ and $\beta=0.1$; $D=0.74$ corresponds to the point near the stability threshold in Fig.~\ref{fig:Stability}. (b) Density-density (solid black) and spin-spin (dashed red) correlation functions at $D=0.74$. There are virtually no density correlations, indicating that the instability is purely spin-wave in character.}
\label{fig:StructureLowField}%
\end{figure}

\begin{figure}[tb]
\includegraphics[width=86mm]{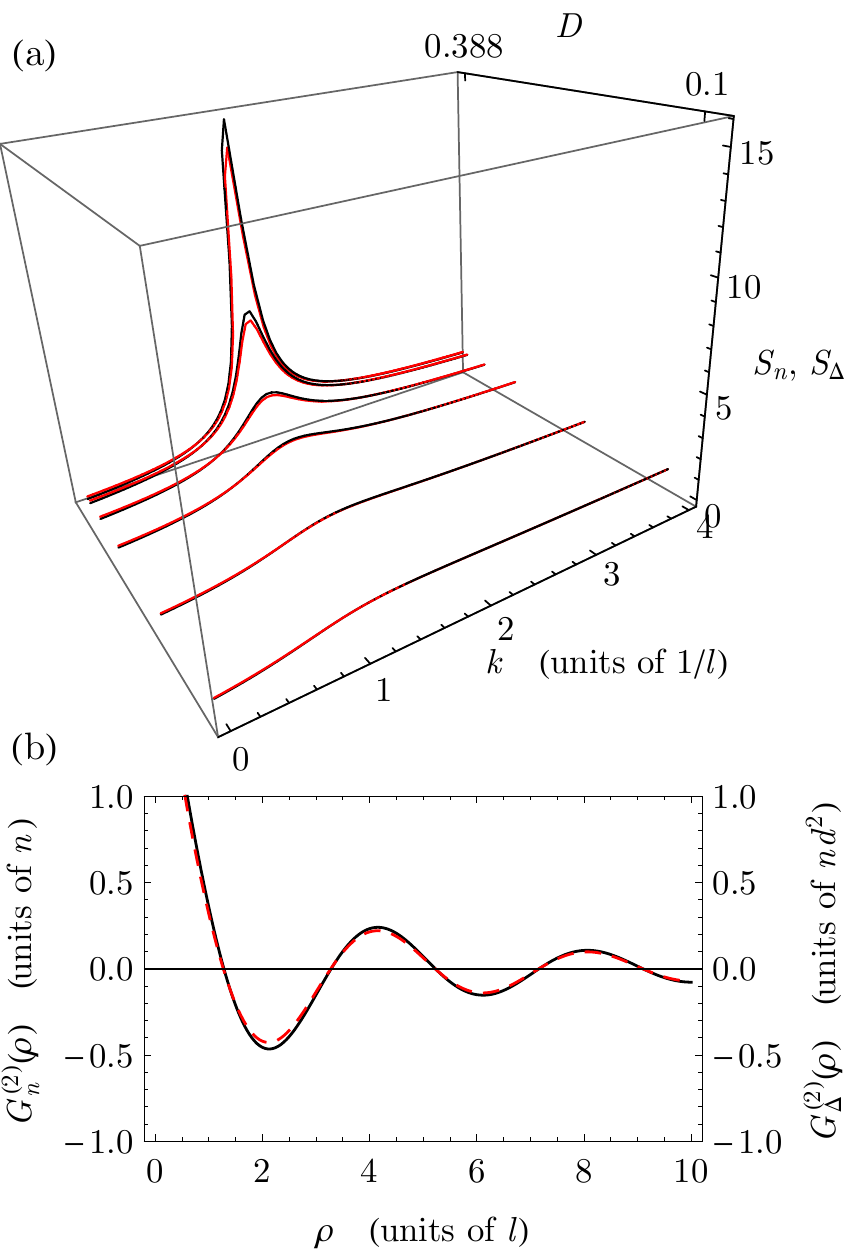}
\caption{(Color online.) (a) Density (black) and spin (red) static structure factors at $\gamma=2$ and $\beta=0.4$; $D=0.388$ corresponds to the point near the stability threshold in Fig.~\ref{fig:Stability}. (b) Density-density (solid black) and spin-spin (dashed red) correlation functions at $D=0.388$. The density and spin behaviors are identical, indicating that the spin fluctuations arise as a trivial consequence of the spin separation cause by the density fluctuations.}
\label{fig:StructureHighField}
\end{figure}

\begin{figure}[tb]
\includegraphics[width=86mm]{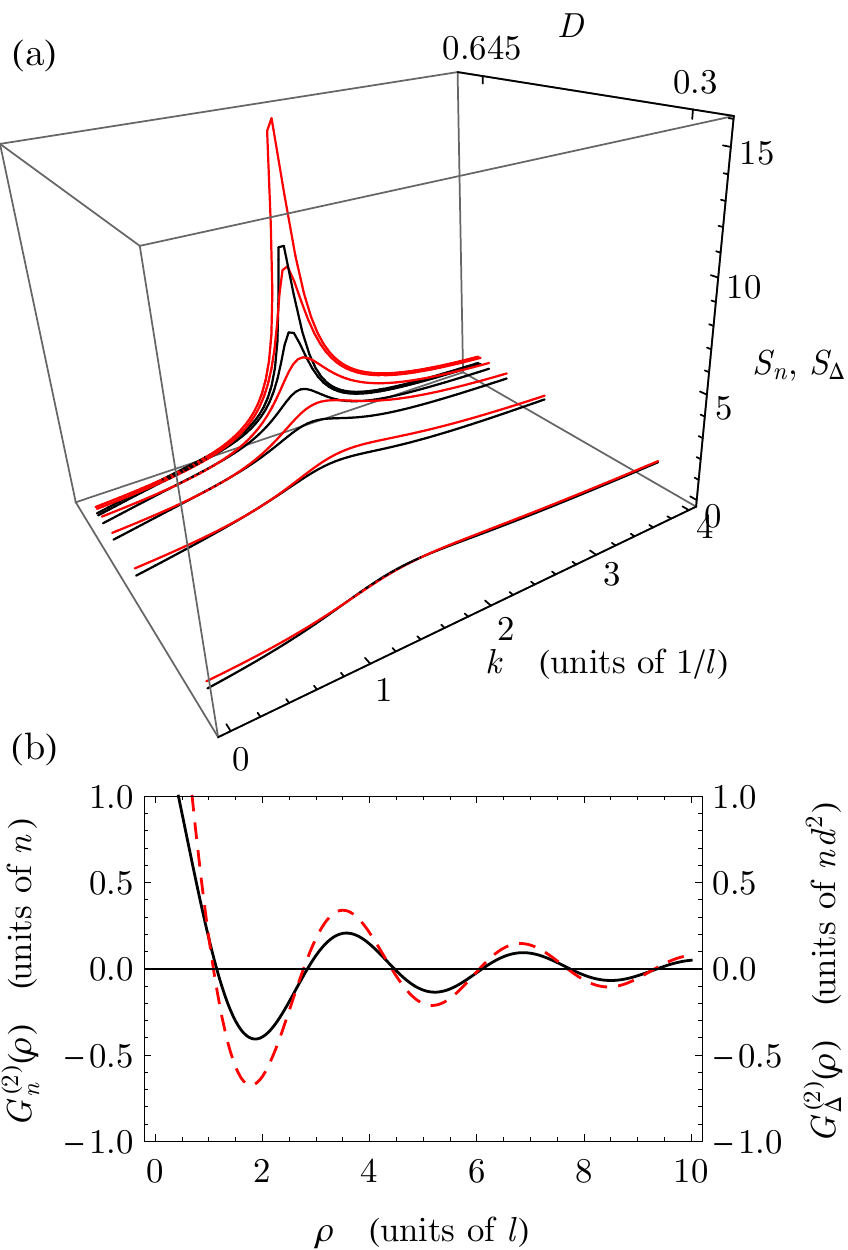}
\caption{(Color online.) (a) Density (black) and spin (red) static structure factors for $\gamma=2$ and $\beta=1.75$; $D=0.645$ correspond to the point near the stability threshold in Fig.~\ref{fig:Stability}. (b) Density-density (solid black) and spin-spin (dashed red) correlation functions at $D=0.645$. In this intermediate regime, both density- and spin-wave behaviors are important.}
\label{fig:StructureMidField}
\end{figure}

The peak in the momentum distribution suggests that the fluctuations arise at a characteristic wavelength. To further explore the nature of the instability, we consider both the response of the system to external perturbations---characterized by the static structure factors $S_{n}(\vec{k})$ and $S_{\Delta}(\vec{k})$--and the spontaneous fluctuations of the ground state---characterized by the two-point correlation functions $G_{n}^{(2)}({\boldsymbol\rho})$ and $G_{\Delta}^{(2)}({\boldsymbol\rho})$. In Figs.~\ref{fig:StructureLowField}a, \ref{fig:StructureHighField}a, and \ref{fig:StructureMidField}a, we have plotted $S_{n}(\vec{k})$ and $S_{\Delta }(\vec{k})$ for a range of values of $D$ that approach the stability threshold in the low-field limit, the cross-over region, and the high-field limit, respectively.  Figures \ref{fig:StructureLowField}b, \ref{fig:StructureHighField}b, and \ref{fig:StructureMidField}b show $G_{n}^{(2)}({\boldsymbol\rho})$ and $G_{\Delta}^{(2)}({\boldsymbol\rho})$ near the instability threshold in each of these regimes.

In the low-field limit (Fig.~\ref{fig:StructureLowField}), we observe a nearly featureless density structure factor, indicating that the system will respond only weakly to fluctuations in the external confining potential. In other words, the system is not susceptible to density-wave fluctuations. On the other hand, in $S_{\Delta}(\vec{k})$ we observe a strong feature appearing at $k\approx2.1/l$ that diverges as $D$ approaches the stability threshold. This indicates that the system is highly susceptible to spin-wave fluctuations and will therefore respond strongly to fluctuations in the external field. In addition, the instability emerges due to the destabilization caused by the onset of spin-wave fluctuations. The onset of spin-waves without the simultaneous onset of density-wave arises as a consequence of the dielectric nature of the system, and hence we dub this a dielectric instability. This discussion is further clarified by analyzing the correlation functions. Apart from a well-known divergence at small $\vec{k}$~\cite{Sykes2012}, the density-density correlation function $G_{n}^{(2)}({\boldsymbol\rho})$ is small for all values of $\vec{k}$, indicating that there are little to no density-density correlations. That is, there are no spontaneous density-wave fluctuations.

In contrast, in the high-field limit (Fig.~\ref{fig:StructureHighField}), we observe that both spin-wave and density-wave fluctuations arise. In the limit of high fields, $S_{n}(\vec{k})$ and $S_{\Delta}(\vec{k})$ are identical, and we see a strong feature in both structure factors appearing at $k\approx1.6/l$. As $D$ approaches the stability threshold, the feature diverges, indicating a destabilization. The correlation functions $G_{n}^{(2)}({\boldsymbol\rho})$ and $G_{\Delta}^{(2)}({\boldsymbol\rho})$ are also identical and indicate that the onset of the spin-wave is a consequence of the onset of the density-wave, in that the gas is fully polarized, and therefore the density fluctuations trivially give rise to polarization fluctuations through the spatial separation of domains of aligned dipoles. At the stability threshold, these density fluctuations are large enough to destabilize the gas, and the gas collapses in the way described before.

Finally, in the cross-over region (Fig.~\ref{fig:StructureMidField}), we observe that the structure factors are featureless far from the stability threshold, but as the stability threshold is approached, strong features again arise in both structure factors at $k\approx1.9/l$. in contrast to both the low-field and high-field limits, in this regime, the density- and spin-wave branches of the dispersion relation are strongly coupled. This indicates that fluctuations in the external field will induce a response in the density, and fluctuations in the external potential will induce a response in the polarization.

\section{Physical Picture of the Instabilities\label{sec:Discussion}}

A full picture of the instabilities seen in this system is as follows.

\begin{figure}[tb]
\includegraphics[width=86mm]{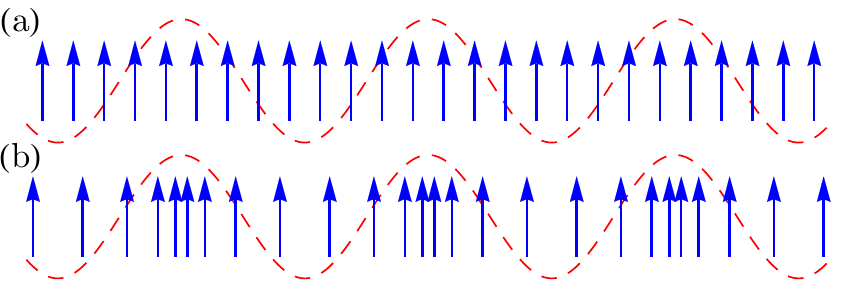}\caption{(Color online.) Schematic of the density-wave instability. (a) The gas is fully polarized with uniform density for small interaction strengths. (b) Density-wave fluctuations arise, leading to local clumping of the gas where it is energetically favorable for the dipoles to line up end to end, despite the energy cost of the transverse harmonic trapping.}
\label{fig:DensityWaveSchematic}
\end{figure}

In a spherical trap, the attractive interaction between molecules lined up end-to-end and the repulsive interaction of laterally aligned dipoles cause the gas to stretch in the axial direction at the expense of an increase in potential energy associated with the trap. At large enough density, this stretching is enough to cause a collapse but can be stabilized by introducing repulsive contact interactions between the dipoles~\cite{Bohn09b}. By contrast, in quasi-2D, the roton mode softens and is mixed into the ground state at finite momentum, leading to density fluctuations. These fluctuations arise at a large wavelength, giving rise to local instabilities analogous to the collective instability occurring in spherically trapped systems~\cite{Bohn09b, Santos03}. In our system, this should occur at large $\beta$, as increasing density will quickly drive the gas toward this density-wave instability. This process is illustrated schematically in Fig.~\ref{fig:DensityWaveSchematic}.

In this high-field limit, the gas is fully polarized. The low-energy excitations above the mean-field ground-state are described by a single dispersion branch that displays a roton-maxon character at large enough interaction strengths (Fig.~\ref{fig:DispersionHighBeta}). The roton minimum appears at finite wavelength because for large enough interaction strengths, it is energetically favorable for the molecules to \emph{locally} pile up end-to-end, despite the energy cost associated with the axial trapping. This leads to a clumping of molecules with a characteristic wavelength of about four times the trap length $l$ (Fig.~\ref{fig:StructureHighField}). Therefore, within one of these clumps, the aspect ratio of the gas is nearly unity, and the clump locally collapses in a manner perfectly analogous to the case of a fully polarized, dipole BEC in a spherically symmetric trap~\cite{Bohn09b}. This picture is verified by means of the two-point, density-density correlation functions, which display strong correlations over large distances ($\sim 10l$) at a wavelength of about $4l$, indicating the presence of a density-wave fluctuation. In addition, the density structure factor develops a strong feature at this wavelength which diverges as the stability threshold is approached ($D\approx0.388$), indicating that the system will respond strongly to small fluctuations in the trapping potential.

\begin{figure}[tb]
\includegraphics[width=86mm]{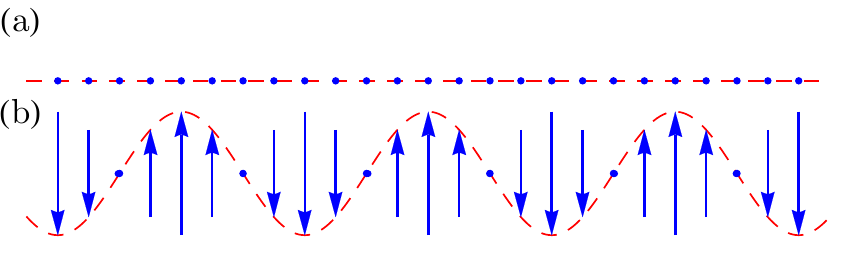}\caption{(Color online.) Schematic of the
spin-wave instability. (a) The system has a uniform density and uniform
zero polarization at small interaction strengths.
(b) For large enough interaction strengths, the system still has a zero
average polarization, but there are domains of oppositely aligned polarization,
giving rise to strong interactions between these domains.
}%
\label{fig:SpinWaveSchematic}%
\end{figure}

In the low-field limit (Fig.~\ref{fig:SpinWaveSchematic}), the system has nearly a zero polarization. Since increasing the interaction strength drives the system toward zero polarization and therefore a perfectly screened mean-field ground state, the system is not susceptible to density-wave fluctuations. The dispersion relations above this mean-field ground state consists of two weakly-coupled modes that cross at large enough interaction strengths (Fig.~\ref{fig:DispersionLowBeta}). One of the modes is a spin-wave mode gapped at zero momentum that undergoes an avoided crossing with the gapless density-wave mode, leading to a roton-like feature at finite momentum that softens as $D$ is increased. The avoided crossing is narrow, which means that the character of the lower branch of the full dispersion relation switches, and the instability that arises therefore has a spin-wave character. Thus, far from the stability threshold, the polarization is uniform, but as the stability threshold is approached, fluctuations arise at a wavelength of about $3l$ that give rise to domains of opposite polarization. This is explained by the fact that at large enough density, the energy cost for flipping domains of dipoles against the external field becomes on the order of the energy benefit from attractive interactions between nearby domains of oppositely-aligned dipoles. Effectively, the local field due to nearby domains of dipoles nearly cancels out the external field, zeroing out the energy cost of aligning dipoles against the field. These fluctuations are of pure spin-wave character, as density-wave fluctuations are suppressed due to the net zero polarization of the mean-field ground state. This is shown in Fig.~\ref{fig:StructureLowField}, where the density-density correlation functions are featureless, whereas the spin-spin correlation functions display strong correlations over large distances. The spin structure factor develops a strong feature at a wavelength of about $3l$, and this feature diverges at the stability threshold. The instability is caused by the strong lateral interactions between adjacent domains of opposite polarization, indicating that the instigated collapse should be longitudinal.

The two instabilities just described are characterized by different competitions of energy scales. The density-wave roton arises due to the competition of trap and interaction energy scales, whereas the spin-wave roton arises due to the competition between the zero-field splitting and the interaction energy scales. In the high-field limit, the gas is fully polarized, and the zero-field splitting is irrelevant, whereas in the low field limit, the influence of the zero-field splitting dominates over the influence of the external field.

\begin{figure}[tb]
\includegraphics[width=86mm]{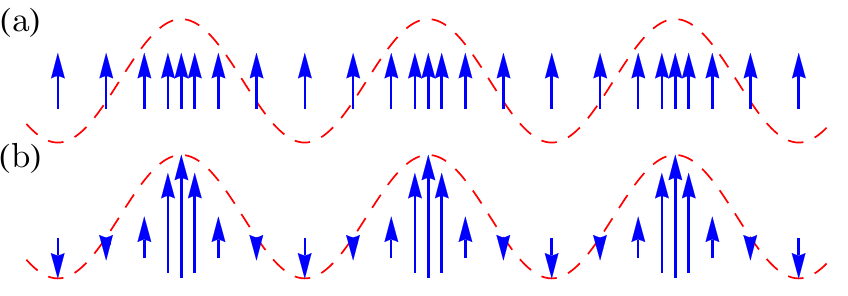}
\caption{(Color online.) Schematic
of the spin-wave-assisted density-wave instability. (a) A density-wave is set
up in which the polarization is essentially uniform. The polarization within
the high-density regions is small enough that the self-attraction within the
domain is too small to destabilize the gas. (b) Spin-wave fluctuations create
neighboring domains of opposing polarizations, leading to a larger density of
aligned spins in the high density regions.}%
\label{fig:AssistedSchematic}%
\end{figure}

Finally, in the intermediate regime, our picture is one of a spin-wave-assisted density-wave instability, as shown schematically in Fig.~\ref{fig:AssistedSchematic}. The argument is as follows. At intermediate fields, a roton-like feature appears in the lower branch of the dispersion relation due to the strong coupling between the spin-wave and density-wave modes (Fig.~\ref{fig:DispersionLowBeta}), although it is not the result of a crossing between the modes. At interaction strengths where this roton appears, strong features appear in both the density and spin structure factors (Fig.~\ref{fig:StructureLowField}). Spontaneous fluctuations in the system can set up a density wave in which the regions of higher density have a polarization small enough that there is no local collapse, indicating that there is no density-wave instability at intermediate field strengths (see dashed curve in Fig.~\ref{fig:Stability}). However, at the same time, the gas is susceptible to spontaneous spin-wave fluctuations with a wavelength comparable to that of the density-wave, leading to larger polarizations within the high-density regions, thereby hastening the local collapse.

\section{Conclusion\label{sec:Conclusion}}

We have identified three distinct mechanisms for the collapse of a two-state, dipolar BEC in quasi-2D under the influence of an external polarizing field. The strong-field behavior is the well-understood density-wave rotonization that occurs in fully polarized dipole BEC's trapped in quasi-2D. It is a consequence of the competition between energy scales of the attractive interactions between dipoles lined up end-to-end and the confinement induced by the trapping. In contrast, the low-field and intermediate-field behaviors are consequences of the dielectric character of the system, where the polarizability of the individual molecules in the condensate plays a crucial role. At very low fields, the gas is stabilized against density-wave fluctuations due to screening in the ground state, and interactions tend to drive the system towards zero polarization. However, the attractive interactions that arise when adjacent domains of opposite polarization appear are strong enough to overcome the energy cost associated with flipping dipoles against the external field. In the intermediate field, the system is stabilized against a density-wave collapse due to the fact that interactions drive the system towards a low polarization, partially screened state. However, superposed polarization and density-waves lead to local polarizations large enough to instigate a collapse analogous to the instability that occurs in the high field limit.

We have also identified three molecules that are candidates for experimentally realizing these results. While gases composed of rigid-rotor molecules such as RbCs are closest to being made quantum degenerate, the densities required for accessing the dielectric instability are experimentally inaccessible. Even if the densities were achievable, it is likely that three-body losses would dominate in those regimes~\cite{Ticknor2010,Wang2011}. For this reason, we focus on the molecules with either a $\Lambda$-doublet or a doublet-$\Sigma$ structure.

The zero-field splitting of the $\Lambda$-doublet in ThO is $\Delta = 30~\textrm{kHz}$ ($\Delta=4.6\times10^{-12}~\tm{au}$)~\cite{Vutha11, Buchachenko2010}. Given a trap frequency of 5 kHz, this corresponds to  $\gamma = 2$, which is what we have used for the majority of the results presented in this paper. The zero-field dielectric instability occurs at $D = 0.75$, and since the maximum dipole moment (in strong applied fields) of ThO is $d_0 = 3.89~\textrm{D}$ ($d_0=1.53~\tm{au}$)~\cite{Vutha11, Buchachenko2010}, this corresponds to a 2D density of $n_{2\tm{D}} = 7.75 \times 10^7/\tm{cm}^2$. The maximum 3D density $n_{2\tm{D}} /\sqrt{\pi}l$ occurs at the center of the trap, and for this choice of parameters, $n_{3\tm{D}} = 8.45\times10^{12}/\tm{cm}^3$. In molecules such as SrF, there are two low-lying states of opposite parity whose splitting can be tuned via an external magnetic field. The splitting is zero at 5370 G and is approximately 100 MHz at $5370\pm40$ G~\cite{Perez-Rios2010}, allowing experimenters to easily access small values of $\gamma$, and therefore small values of $D$ and reasonable values of $n$. The combination of a tunable zero-field splitting and the fact that a gas of SrF molecules can be laser-cooled~\cite{Shuman2010, Barry2012} makes SrF an enticing candidate for future experimental investigation of the results presented in this paper.

In real systems, the gases are confined in all three dimensions in a so-called ``pancake'' geometry. In these setups, a low-momentum cutoff is introduced due to the finite size of the system, and the correlations seen in the density-density, two-point correlation functions manifest as density striping~\cite{Ticknor11, Wilson12b, Baillie2015a}. In the low-field limit of our model, we expect polarization striping to appear in the absence of density striping when the gas is harmonically confined in the plane, which is the subject of ongoing investigation.

The theoretical methods presented in the first part of this paper comprise a general method for investigating the mean-field ground state and low-lying excitations of a BEC of dipolar molecules purely analytically. The long-range nature of the dipole-dipole interactions result in momentum-dependent coupling constants in the Bogoliubov-de Gennes fluctuation Hamiltonian, but this Hamiltonian can be diagonalized analytically at every non-zero momentum with a sequence of physically motivated canonical transformations. This procedure generalizes to situations where the internal states of the molecules in a BEC are linearly coupled, and the molecules interact via direct, long-range interactions.

We acknowledge many fruitful discussions with J.~L.~Bohn, B.~L.~Johnson, R.~V.~Krems, C.~Ticknor, E.~Timmermans, and H.~R.~Sadeghpour. This work was partially funded by NSF Grant Nos.~PHY-1516337 and PHY-1516421.

\appendix

\section{Interaction Parameters\label{app:InteractionParameters}}

In this appendix, we outline the derivation showing that the interaction
parameter,%
\begin{align}
\lambda_{\kappa\kappa^{\prime}}\left(  \vec{k},\vec{k}^{\prime}\right)   &
=\frac{d_{\kappa^{\prime}}d_{\kappa}}{l^{2}}\int d^{3}x\left\vert f_{\kappa
}\left(  z\right)  \right\vert ^{2}\frac{e^{-i\vec{k}\cdot{\boldsymbol\rho}}%
}{\sqrt{A}}\nonumber\\
&  \quad\mbox{}\times\int d^{3}x^{\prime}V\left(  \vec{x}-\vec{x}^{\prime
}\right)  \left\vert f_{\kappa^{\prime}}\left(  z^{\prime}\right)  \right\vert
^{2}\frac{e^{i\vec{k}^{\prime}\cdot{\boldsymbol\rho}^{\prime}}}{\sqrt{A}},
\end{align}
where%
\begin{equation}
V\left(  \vec{r}\right)  =\frac{1-3\left(  \hat{\mathbf{n}}\cdot
\hat{\mathbf{r}}\right)  ^{2}}{r^{3}},
\end{equation}
can be written as%
\begin{equation}
\lambda_{\kappa\kappa^{\prime}}\left(  \vec{k},\vec{k}^{\prime}\right)
=\delta_{\vec{k},\vec{k}^{\prime}}\lambda_{\vec{k},\kappa\kappa^{\prime}},
\end{equation}
where
\begin{subequations}
\label{eqn:InteractionParameterGeneral}%
\begin{align}
\lambda_{\vec{k},\kappa\kappa^{\prime}}
&=
\frac{d_{\kappa}d_{\kappa^{\prime}}}{l^{2}}4\pi\left(  n_{z}^{2}-\frac{1}{3}\right)  \int dz\left\vert
f_{\kappa}\left(  z\right)  \right\vert ^{2}\left\vert f_{\kappa^{\prime}%
}\left(  z\right)  \right\vert ^{2}\nonumber\\
&  \quad\mbox{}-\pi\frac{d_{\kappa}d_{\kappa^{\prime}}}{l^{2}}\left(
n_{z}-i\left(  \hat{\mathbf{n}}\cdot\hat{\mathbf{k}}\right)  \right)
^{2}kF_{\kappa\kappa^{\prime}}\left(  k\right) \nonumber\\
&  \quad\mbox{}-\pi\frac{d_{\kappa}d_{\kappa^{\prime}}}{l^{2}}\left(
n_{z}+i\left(  \hat{\mathbf{n}}\cdot\hat{\mathbf{k}}\right)  \right)
^{2}kF_{\kappa^{\prime}\kappa}\left(  k\right)  ,\\
F_{\kappa\kappa^{\prime}}\left(  k\right)
&=
\int_{0}^{\infty}due^{-ku}
\int_{-\infty}^{\infty} du^{\prime}
\nonumber\\
&\quad\mbox{}\times
\left\vert
f_{\kappa}\left(\frac{u^{\prime}+u}{2}\right)
f_{\kappa^{\prime}}\left(  \frac{u^{\prime}-u}{2}\right)
\right\vert^{2}.
\end{align}
When a Gaussian ansatz for the axial wave functions is employed, given by%
\end{subequations}
\begin{equation}
f_{\kappa}\left(  z\right)  =\frac{1}{\sqrt{l_{\kappa}\sqrt{\pi}}}%
e^{-z^{2}/2l_{\kappa}^{2}},
\end{equation}
the integrals in Eq.~(\ref{eqn:InteractionParameterGeneral}) can be readily
computed, and the result is%
\begin{align}
\lambda_{\vec{k},\kappa\kappa^{\prime}}  &  =\lambda_{\kappa\kappa^{\prime}%
}F\left(  \frac{l_{\kappa\kappa^{\prime}}}{\sqrt{2}}\vec{k}\right)  ,\\
\lambda_{\kappa\kappa^{\prime}}  &  =\frac{4\sqrt{2\pi}}{3}\frac{d_{\kappa
}d_{\kappa^{\prime}}}{l^{3}}\frac{l}{l_{\kappa\kappa^{\prime}}}\frac
{3n_{z}^{2}-1}{2},\\
F\left(  \vec{x}\right)  &=
1-\frac{3}{2}\sqrt{\pi}\frac{2n_{z}^{2}-2\left(\hat{\mathbf{n}}\cdot\hat{\mathbf{x}}\right) ^{2}}{3n_{z}^{2}-1}xe^{x^{2}%
}\mathrm{erfc}\left(  x\right)  ,
\end{align}
where erfc is the complementary error function, and
\begin{align}
l_{\kappa\kappa^{\prime}} = \sqrt{\frac{l_{\kappa}^2+l_{\kappa^{\prime}}^2}{2}}.
\end{align}

We start by transforming to center-of-mass-like coordinates,
\begin{align}
\vec{x}  &  =\frac{\vec{R}+\vec{r}}{2},\\
\vec{x}^{\prime}  &  =\frac{\vec{R}-\vec{r}}{2},
\end{align}
in which case%
\begin{align}
\lambda_{\kappa\kappa^{\prime}}\left(  \vec{k},\vec{k}^{\prime}\right)
&=
\frac{d_{\kappa^{\prime}}d_{\kappa}}{l^{2}}\frac{1}{8}\int dz\int dz^{\prime}
\nonumber\\
& \quad\mbox{}\times
\left\vert f_{\kappa}\left(  \frac{z+z^{\prime}}{2}\right)\right\vert^{2}\left\vert f_{\kappa^{\prime}}\left(  \frac{z^{\prime}-z}{2}\right) \right\vert ^{2}
\nonumber\\
&  \quad\mbox{}\times\int d^{2}\rho\frac{e^{-i\vec{k}\cdot{\boldsymbol\rho}%
/2}}{\sqrt{A}}V\left(  \vec{r}\right)  \frac{e^{-i\vec{k}^{\prime}%
\cdot{\boldsymbol\rho}/2}}{\sqrt{A}}
\nonumber\\
&  \quad\mbox{}\times\int d^{2}\rho^{\prime}\frac{e^{-i\vec{k}\cdot
{\boldsymbol\rho}^{\prime}/2}}{\sqrt{A}}\frac{e^{i\vec{k}^{\prime}%
\cdot{\boldsymbol\rho}^{\prime}/2}}{\sqrt{A}},
\end{align}
where%
\begin{align}
\vec{r}  &  =\left(  {\boldsymbol\rho},z\right)  ,\\
\vec{R}  &  =\left(  {\boldsymbol\rho}^{\prime},z^{\prime}\right)  .
\end{align}
The last integral evaluates to $4\delta_{\vec{k},\vec{k}^{\prime}}$, in which
case we define%
\begin{align}
\lambda_{\vec{k},\kappa\kappa^{\prime}}
&=
\frac{d_{\kappa^{\prime}}d_{\kappa}}{l^{2}}\frac{1}{2}\int dz\int dz^{\prime}
\nonumber\\
& \quad\mbox{}\times\left\vert f_{\kappa
}\left(  \frac{z+z^{\prime}}{2}\right)  \right\vert ^{2}\left\vert
f_{\kappa^{\prime}}\left(  \frac{z^{\prime}-z}{2}\right)  \right\vert ^{2}\\
&  \quad\mbox{}\times\int d^{2}\rho\frac{e^{-i\vec{k}\cdot{\boldsymbol\rho}}%
}{A}V\left(  \vec{r}\right)  .
\end{align}

The next step involves three convenient identities involving the dipole-dipole
interaction. The first gives the interaction as the sum of a short-range and a
long range piece, given by~\cite{Rosenkranz2013}%
\begin{equation}
V\left(  \vec{r}\right)  =\frac{1-3\left(  \hat{\mathbf{n}}\cdot\hat{\vec{r}%
}\right)  ^{2}}{r^{3}}=-\frac{4\pi}{3}\delta\left(  \vec{r}\right)  -\left(
\hat{\mathbf{n}}\cdot\vec{{\boldsymbol\nabla}}_{\vec{r}}\right)  ^{2}\frac
{1}{r}.
\end{equation}
The second identity,%
\begin{align}
\left(  \hat{\mathbf{n}}\cdot\vec{{\boldsymbol\nabla}}_{\vec{r}}\right)  ^{2}
&  =\left(  \hat{\mathbf{n}}\cdot\left(  \hat{\mathbf{z}}\frac{\partial
}{\partial z}+\vec{{\boldsymbol\nabla}}_{{\boldsymbol\rho}}\right)  \right)
^{2}\nonumber\\
&  =n_{z}^{2}\frac{\partial^{2}}{\partial z^{2}}+2n_{z}\frac{\partial
}{\partial z}\left(  \hat{\mathbf{n}}\cdot\vec{{\boldsymbol\nabla}%
}_{{\boldsymbol\rho}}\right)  +\left(  \hat{\mathbf{n}}\cdot\vec
{{\boldsymbol\nabla}}_{{\boldsymbol\rho}}\right)  ^{2},
\end{align}
allows us to explicitly separate the axial and transverse dependence of the
interaction. Finally, making use of the fact that%
\begin{equation}
\frac{\partial^{2}}{\partial z^{2}}={\nabla}_{\vec{r}}^{2}-{\nabla}
_{{\boldsymbol\rho}}^{2},
\end{equation}
we can write%
\begin{equation}
\frac{\partial^{2}}{\partial z^{2}}\frac{1}{r}=-4\pi\delta\left(  \vec
{r}\right)  -{\nabla}_{{\boldsymbol\rho}}^{2}\frac{1}{r},
\end{equation}
in which case
\begin{align}
\frac{1-3\left(  \hat{\mathbf{n}}\cdot\hat{\mathbf{r}}\right)  ^{2}}{r^{3}}
&=
\frac{8\pi}{3}\frac{3n_{z}^{2}-1}{2}\delta\left(  \vec{r}\right)
\nonumber\\
&\quad\mbox{}
+\left(n_{z}^{2}{\nabla}_{{\boldsymbol\rho}}^{2}-\left(  \hat{\mathbf{n}}\cdot
\vec{{\boldsymbol\nabla}}_{{\boldsymbol\rho}}\right)  ^{2}\right)  \frac{1}%
{r}\nonumber\\
&  \quad\mbox{}-2n_{z}\frac{\partial}{\partial z}\left(  \hat{\mathbf{n}}%
\cdot\vec{{\boldsymbol\nabla}}_{{\boldsymbol\rho}}\right)  \frac{1}{r}.
\end{align}
Using properties of the Fourier transform and the identities%
\begin{align}
J_{0}\left(  kR\right)   &  =\int_{0}^{2\pi}d\phi\frac{e^{-ikR\cos\phi}}{2\pi
},\\
\frac{e^{-k\left\vert z\right\vert }}{k}  &  =\int_{0}^{\infty}d\rho\frac{\rho
J_{0}\left(  k\rho\right)  }{\sqrt{z^{2}+\rho^{2}}},
\end{align}
the interaction parameter can be written as%
\begin{align}
\lambda_{\vec{k},\kappa\kappa^{\prime}}  &  = \frac{d_{\kappa^{\prime}%
}d_{\kappa}}{l^{2}}\frac{8\pi}{3}\frac{3n_{z}^{2}-1}{2}\int dz\left\vert
f_{\kappa}\left(  z\right)  \right\vert ^{2}\left\vert f_{\kappa^{\prime}%
}\left(  z\right)  \right\vert ^{2}\nonumber\\
&  \quad\mbox{}-\frac{d_{\kappa}d_{\kappa^{\prime}}}{l^{2}}\pi\left(
n_{z}-i\left(  \hat{\mathbf{n}}\cdot\hat{\mathbf{k}}\right)  \right)
^{2}kF_{\kappa\kappa^{\prime}}\left(  k\right) \nonumber\\
&  \quad\mbox{}-\frac{d_{\kappa^{\prime}}d_{\kappa}}{l^{2}}\pi\left(
n_{z}+i\left(  \hat{\mathbf{n}}\cdot\hat{\mathbf{k}}\right)  \right)
^{2}kF_{\kappa^{\prime}\kappa}\left(  k\right)  ,
\end{align}
where
\begin{align}
F_{\kappa\kappa^{\prime}}\left(  k\right)
&=
\int_{0}^{\infty}dze^{-kz}\int_{-\infty}^{\infty}dz^{\prime}
\nonumber\\
&\quad\mbox{}\times
\left\vert f_{\kappa}\left(  \frac{z^{\prime}+z}{2}\right)
f_{\kappa^{\prime}}\left(  \frac{z^{\prime}-z}{2}\right)  \right\vert ^{2}.
\end{align}
We note that%
\begin{equation}
\lambda_{\vec{k},\kappa\kappa^{\prime}}=\lambda_{\vec{k},\kappa^{\prime}%
\kappa}^{\ast}=\lambda_{-\vec{k},\kappa^{\prime}\kappa},
\end{equation}
which implies that%
\begin{equation}
\Lambda_{\vec{k},\sigma\sigma^{\prime}}=\Lambda_{\vec{k},\sigma^{\prime}%
\sigma}^{\ast}=\Lambda_{-\vec{k},\sigma^{\prime}\sigma},
\end{equation}
by way of Eq.~\ref{eqn:DressedInteractionCouplings}. In the case where the
external field is transverse, i.e.~parallel to the trap axis,%
\begin{align}
n_{z}  &  =1,\\
\hat{\mathbf{n}}\cdot\hat{\mathbf{k}}  &  =0,
\end{align}
in which case
\begin{align}
\lambda_{\vec{k},\kappa\kappa^{\prime}}
&=
\frac{d_{\kappa^{\prime}}d_{\kappa}}{l^{2}}\frac{8\pi}{3}\int dz\left\vert f_{\kappa}\left(  z\right)
\right\vert ^{2}
\left\vert f_{\kappa^{\prime}}\left(  z\right)  \right\vert^{2}
\nonumber\\
&\quad\mbox{}-
\frac{d_{\kappa}d_{\kappa^{\prime}}}{l^{2}}2\pi k\int dzdz^{\prime}e^{-k\left\vert z-z^{\prime}\right\vert }
\nonumber\\
&\quad\mbox{}\times
\left\vert f_{\kappa}\left(  z\right)\right\vert^{2}
\left\vert f_{\kappa^{\prime}}\left(  z^{\prime}\right)  \right\vert^{2}.
\end{align}

\section{Linear response theory\label{app:LinearResponseTheory}}

In this appendix, we outline the specific definitions used for the response functions~\cite{Zambelli2000}. If the many-body system is subject to a perturbing Hamiltonian of the form,
\begin{equation}
\delta\hat{H}\left(  t\right)  =-\int d^{2}\rho\hat{A}\left(  {\boldsymbol\rho},t\right)  f\left(  {\boldsymbol\rho},t\right)  ,
\end{equation}
then, according to linear response theory, the response of the observable $\hat{B}$ to this perturbation is given by
\begin{equation}
\delta\langle\hat{B}\left(  {\boldsymbol\rho},t\right)  \rangle=\int_{-\infty
}^{\infty}dt^{\prime}\int d^{2}\rho^{\prime}\chi_{BA}\left(  {\boldsymbol\rho
},t;{\boldsymbol\rho}^{\prime},t^{\prime}\right)  f\left(  {\boldsymbol\rho
}^{\prime},t^{\prime}\right)  ,
\end{equation}
where%
\begin{equation}
\chi_{BA}\left(  {\boldsymbol\rho},t;{\boldsymbol\rho}^{\prime},t^{\prime
}\right)  =\frac{i}{\hbar}\langle[ \hat{B}\left(  {\boldsymbol\rho},t\right)
,\hat{A}\left(  {\boldsymbol\rho}^{\prime},t^{\prime}\right)  ] \rangle
\Theta\left(  t-t^{\prime}\right)  ,
\end{equation}
is called the response function, or susceptibility, and the expectation value
is taken in the quasi-particle vacuum. This system is both time-translation
and space-translation invariant, in which case we can replace the response
function with%
\begin{equation}
\chi_{BA}\left(  \boldsymbol\rho,t\right)  =\chi_{BA}\left(  {\boldsymbol\rho},t;0,0\right)  .
\end{equation}
We are interested in the static parts of these susceptibilities, given by%
\begin{align}
\tilde{\chi}_{BA}\left(  \vec{k}\right)   &  =\int d\omega\int_{-\infty
}^{\infty}dte^{i\omega t}\int\frac{d^{2}\rho}{2\pi}e^{-\vec{k}\cdot
{\boldsymbol\rho}}\chi_{BA}\left(  {\boldsymbol\rho},t\right)  .
\end{align}

Specifically, we are interested in the response of the polarization
$\hat{\Delta}$ to small changes in the external electric field and the
response of the density $\hat{n}$ to external perturbations in the trapping
potential. Thus, we are interested in the cases where $\hat{\Delta}=\hat
{B}=\hat{A}$ and $\hat{n}=\hat{B}=\hat{A}$. It can be shown that the polarization and density
response functions can be written as
\begin{align}
\tilde{\chi}_{n}\left(  \vec{k}\right)   &  =i\frac{n}{4\hbar}\left(
\frac{\delta\bar{n}}{n}+S_{n}\left(  \vec{k}\right)  \right)  ,\\
\tilde{\chi}_{\Delta}\left(  \vec{k}\right)   &  =i\frac{nd^{2}}{4\hbar
}\left(  \frac{\delta\bar{n}}{n}+S_{\Delta}\left(  \vec{k}\right)  \right)  .
\end{align}
Since the depletions must satisfy $\delta\bar{n}\ll n$, it is apparent that
the susceptibilities are characterized completely by the structure factors
$S_{n}\left(  \vec{k}\right)  $ and $S_{\Delta}\left(  \vec{k}\right)  $.

\bibliographystyle{apsrev}

\end{document}